# An Irreversible Synthetic Route to an Ultra-Strong Two-Dimensional Polymer

Yuwen Zeng[1], Pavlo Gordiichuk[1], Takeo Ichihara[1], Ge Zhang[1], Xun Gong[1], Sandoz-Rosado Emil[2], Eric D. Wetzel[2], Jason Tresback[3], Jing Yang[1], Zhongyue Yang[1], Daichi Kozawa[1], Matthias Kuehne[1], Pingwei Liu[1], Albert Tianxiang Liu[1], Jingfan Yang[1], Heather J. Kulik[1], Michael S. Strano[1]*

[1]Department of Chemical Engineering, Massachusetts Institute of Technology, Cambridge, MA 02139, USA

[2]U.S. Army Research Laboratory, Aberdeen Proving Ground, MD 21005-5069, USA

[3]Center for Nanoscale Systems, Harvard University, Cambridge, MA 02139, USA

**Polymers that extend covalently in two dimensions have attracted recent attention[1,2] as a means of combining the mechanical strength and in-plane energy conduction of conventional two-dimensional (2D) materials[3,4] with the low densities, synthetic processability, and organic composition of their one-dimensional counterparts.[5,6] Efforts to date have proven successful in forms that do not allow full realization of these properties, such as polymerization at flat interfaces[7,8] or fixation of monomers in immobilized lattices.[9] A frequently employed synthetic approach is to introduce microscopic reversibility, at the cost of bond stability, to achieve 2D crystals after extensive error correction.[10,11] Herein we demonstrate a synthetic route to 2D irreversible polycondensation directly in the solution phase, resulting in covalently bonded 2D polymer platelets that are chemically stable and highly processable. Further fabrication offers highly oriented, free-standing films which exhibit exceptional 2D elastic modulus and yield strength at 50.9 ± 15.0 GPa and 0.976 ± 0.113 GPa, respectively. Platelet alignment is evidenced by a polarized photoluminescence centered at 580 and 680 nm from different dipole transitions. This new synthetic route provides opportunities for 2D polymers in applications ranging from composite structures to molecular sieving membranes.**

Two-dimensional polymers have long been conceptualized and attempts as early as 1935 explored one-dimensional concatenation of amphiphiles confined to the air-water interface.[12,13] This work by Gee and co-workers inspired recent advances towards 2D polymers utilizing surface templates in the work of Stupp[7] and Ozaki[14], promising strategies that might allow the release of such materials from the template in a scalable fashion.[9] Similarly, Cote and co-workers first synthesized two covalent organic frameworks (COF) – which entail reversible solvothermal crystallization – that showed layered unit cells stacked in 3D.[15] This and variants, however, have proven difficult to exfoliate or reprocess into engineering materials with useful properties. Reversible synthetic approaches appear to yield materials with limited chemical and mechanical stability such that exfoliation or isolation is difficult.[10,16] A substantial advance came about recently by utilizing the irreversible nucleophilic aromatic substitutions for 2D polycondensation under solvothermal conditions.[17,18] The resulting bulk powders, unlike previous examples, are chemically stable – pointing to the importance of irreversible bonding in approaching the properties of 1D polymer systems. In contrast, Chemical Vapor Deposition (CVD) has enabled access to a substantial number of 2D crystalline materials such as graphene, dichalcogenides, and hexagonal boron nitride,[19] with extra-ordinary in-plane mechanical properties approaching that of diamond.[3,20] The high temperatures typically needed for CVD exclude polymerization of organic molecules, and hence, 2D organic analogs of 1D polymers have thus far remained elusive. The totality of this prior work points to the importance of irreversible chemical routes, necessarily outside of solvothermal or CVD synthetic methods, with in-plane bonding such that solution-phase exfoliation is possible.

There are several challenges to 2D polymerization directly in bulk solution, but the most fundamental is that for a 2D disk molecule, the number of perimeter addition sites scales with the number of incorporated monomers *i* to the 1/2 power (Fig. S1). For the 3D spherical counterpart, however, it grows much faster with 2/3 power (Fig. S1). This means that as soon as a polymerizing molecular disk grows a defect branch out of the plane, the 3D structure will extend much faster than the desired in-plane 2D disk. Such out-of-plane branches occur easily, with just a single bond rotation of an attached monomer. It is clear that 2D polymerization must fundamentally overcome the high entropic cost of maintaining in-plane bonding. Worst still, irreversible synthetic routes necessarily magnify the impact of single out-of-plane defects with no means of error-correcting.[2]

Our strategy in designing a synthetic approach to overcome these challenges is multifold, and involves an amide condensation of C3-symmetric acid chloride and melamine (Fig. 1a). Our hypothesis is that a strong amide-aromatic conjugation inhibits out-of-plane rotation, meanwhile, the interlayer hydrogen bonding or van der Waals attraction can allow growing disks to absorb monomers from the solution phase and auto-template them onto 2D surfaces, facilitating the 2D growth pathway (Fig. S2). Indeed, chemical kinetic modeling shows that it is possible to produce 2D polymers in excess of 90% yield with a combination of a moderate in-plane to out-of-plane probability ratio ($\gamma$), expected from amide conjugation, along with a relative rate constant acceleration ($\beta$) from monomers adsorbed onto existing 2D platelets as a form of auto-catalytic self-templating (Fig. S3). In our designed reaction system, the inert amide linkage ensures superb mechanical and chemical stability, allowing sonication, harsh acid or heat treatment (see SI for detail). Additionally, triazine cores are intentionally introduced into the structure (Fig 1a) and offer a high density of Lewis bases, leading to protonation in strong acid and thus excellent solubility in such solvents for high processability.

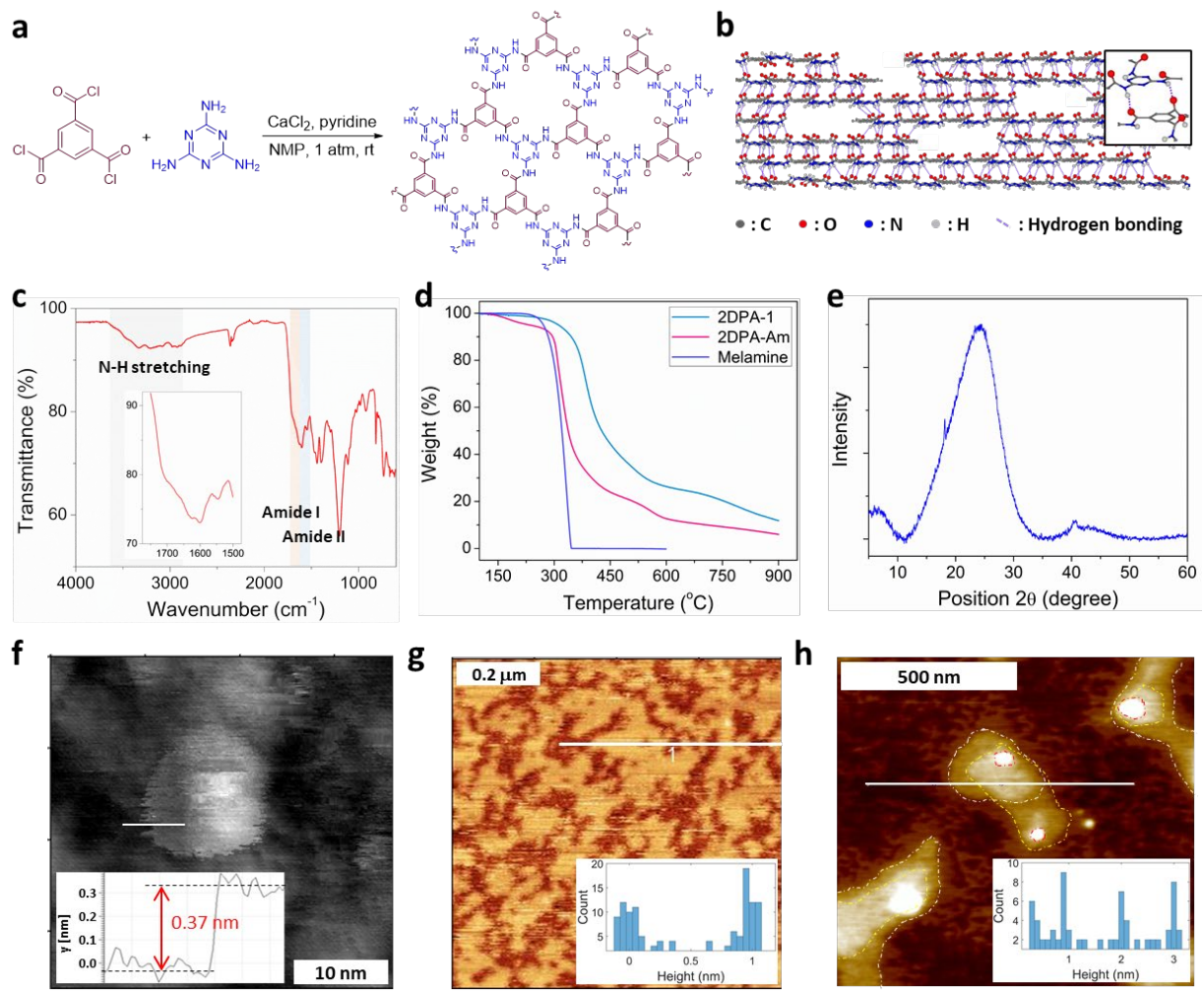

**Figure 1| Synthesis and characterization of a two-dimensional (2D) polymer. a,** Synthetic route to 2D polyaramid, termed **2DPA-1**. **b,** Cross-sectional view of a proposed hydrogen-bonded, interlocked layered structure. A close-up of interlayer hydrogen bonds is showing in the inset. **c,** Fourier-transform infrared (FT-IR) spectroscopy of **2DPA-1**. **d,** Thermal gravimetric analysis (TGA) of **2DPA-1**, **2DPA-Am** (amorphous counterpart of **2DPA-1**, obtained when trimesoyl chloride is replaced by isothaloyl chloride under standard conditions), and melamine. **e,** Powder X-ray diffraction (PXRD) of **2DPA-1**. **f,** High-resolution atomic force microscopy (HR-AFM) image of one individual molecule absorbed on mica and its height profile along the white line (inset). **g,** AFM image of bilayer nanoclusters and its height histogram along the white line (inset). **h,** AFM image of stacked nanosheets; inset shows the height histogram along the white line.

We indeed find that this irreversible 2D polyaramid chemistry enables the monomers to condense in the solution phase under ambient and neutral conditions producing high yields of in-plane bonded 2D polymers, termed **2DPA-1** (Fig. 1a). The Fourier-transform infrared spectrum (FT-IR) confirms the formation of the amide linkages in both amide **I** (1600-1700 cm$^{-1}$) and amide **II** (1500-1600 cm$^{-1}$) regions (Fig. 1c).[21] The broadened peaks suggest that the amides exist in heterogeneous chemical environments. Multiple broad and red-shifted peaks between 2800 and 3500 cm$^{-1}$ are assigned to highly hydrogen-bonded N-H stretching.[22] This observation suggests that the condensed material contains amides that tilt out of the plane, forming interlayer hydrogen bonds.[23] By thermal gravimetric analysis (TGA), **2DPA-1** shows a sharp decomposition curve starting at 312°C, corresponding to 5% weight loss, well above the melamine precursor at 271°C and the amorphous counterpart **2DPA-Am** at 233°C (Fig. 1d). The elevated thermal stability is consistent with both 2D concatenation in-plane and hydrogen bonding out-of-plane. A subsequent differential thermogravimetric (DTG) analysis indicates a very high 2D percentage exceeding 95% (Fig. S9). The powder x-ray diffraction (PXRD) of **2DPA-1** shows an imperfect crystalline structure with a peak centered at about 24.3 degrees (Fig. 1e), corresponding to an average interlayer spacing of 3.66 Å. The lack of long-range ordering of the as-produced **2DPA-1** powder suggests small, isotropically oriented but stacked nanoplatelets (Fig. 1b). This molecular picture agrees with high resolution atomic force microscopy (HR-AFM) after solution deposition. We find that by tuning the deposition conditions, individual platelets (Fig. 1f), bilayer clusters (Fig. 1g), and continuous nanosheets (Fig. 1h) are all observed. The average single layer thickness is found to be 4.04±0.48 Å (Fig. S15) with a bilayer thickness of 7.95±0.72 Å (Fig. S16), both of which approximately agree with the interlayer spacing from PXRD (3.66 Å) and Grazing-incidence wide-angle X-ray scattering (GIWAXS) measurements (3.72 Å) as discussed below. Lastly, densely packed layered structures are observed in Transmission Electron Microscopy (TEM) at the edge of liquid exfoliated powders (Fig. S19). We note that the interlayer thickness of **2DPA-1** is larger than that of typical van der Waals 2D materials such as graphene[24] (3.35 Å) or hBN[25] (3.33 Å). This appears to support the influence of amides rotated out-of-plane allowing interlayer hydrogen bonding in agreement with our molecular modeling (Fig. S35).

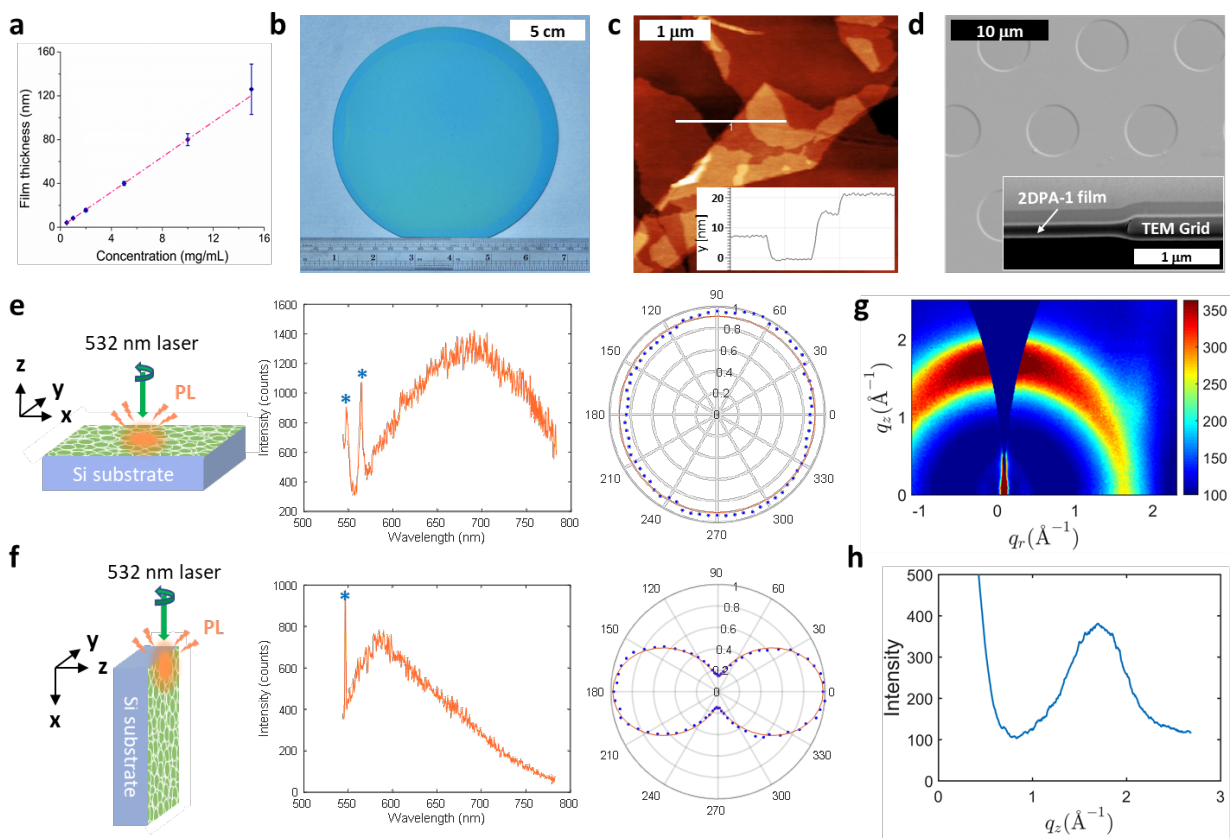

**Figure 2| Characterization of 2DPA-1 nanofilms. a,** The thickness versus concentration dependence of spin-coated **2DPA-1** films on SiO$_2$ covered (300 nm) silicon wafers. Confidence intervals are shown as error bars. **b,** A 6-inch transferred **2DPA-1** film

on a SiO$_2$/Si wafer. **c,** Atomic force microscopy (AFM) image of a transferred **2DPA-1** nanofilm at film edges near which cracks, wrinkles, and folds are observed. The inset shows the height profile along the white line. **d,** Scanning electron microscopy (SEM) images of a suspended **2DPA-1** film on a Si$_3$N$_4$ TEM Grid. Inset: a cross-sectional view of a hole after focused ion beam (FIB) cutting. **e,** Top view photoluminescence (PL) measurement of a **2DPA-1** nanofilm at 532 nm excitation. Left: schematic illustration; middle: PL spectrum, where Si Raman peaks at 550 and 570 nm are labeled with "*"; right: polar plot of the top view. Red fitting curve: intensity = 0.91. **f,** Side view photoluminescence (PL) measurement of a **2DPA-1** nanofilm at 532 nm excitation. Left: schematic illustration; middle: PL spectrum, where the Si Raman peak at 550 nm is labeled with "*"; right: polar plot of the side view. Red fitting curve: intensity = 0.1452 + 0.8365*cos2θ. **g,** Grazing-incidence wide-angle X-ray scattering (GIWAXS) 2D image, and its 1D intensity profile (**h**) near q$_r$ = 0 A$^{-1}$ of the **2DPA-1** film.

To realize compelling applications of 2D materials, it is important that they be processable into 2D films, such as membranes[26], mechanical composites[27], and barrier coatings.[28] However, polycrystallinity and poor alignment often confound this goal.[29-31] For **2DPA-1**, its strong aggregation tendency allows uniform and continuous nanofilms to be easily generated by simple spin-coating a TFA solution (0.5-15 mg/mL) onto flat SiO$_2$, mica, or polycarbonate. The thickness can be well controlled by tuning the solution concentration (Fig. 2a). It is noteworthy that the film thickness can go less than 4 nm, indicating that even a few molecular layers bind strongly enough to form an infinitely extended film (Fig. S20). All measured films have an apparent RMS (root mean square) roughness around 500 pm over a 5x5 μm area (Fig. S23), corresponding to a height variation of no more than 4 molecular layers. We thus conclude that the resulting **2DPA-1** spin-coated films increase in orientation order with platelets aligning in-plane (Fig. 1b). We further developed a film transfer method for rough or even holey substrates (Fig. S42). With the assistance of a polycarbonate transfer layer, a 6-inch, 7-nm thick **2DPA-1** nanofilm was transferred fully intact onto a SiO$_2$/Si wafer (Fig. 2b & 2c and Fig. S44). Despite cracks, wrinkles, and folds are observed at the film edges, the transferred material is flat and continuous. Remarkably, free-standing membranes can even form from the solution phase after drop-casting across empty 5 μm holes, exhibiting excellent formability. SEM top view and cross-sectional views after focused ion beam (FIB) cutting (Fig. 2d) show an absence of scattering defects caused by polycrystalline abrasions or imperfections, supporting the observation of high orientation order within the film.

We find that when excited at 532 nm with polarized laser excitation, the **2DPA-1** in powder form exhibits broad visible fluorescence at 580 nm with isotropic polarization (Fig. S31 and S32). Meanwhile, the spin-coated nanofilm exhibits a different fluorescent emission at 680 nm when excitation orthogonal to the film surface, also with isotropic polarization (Fig. 2e). However, when excitation is parallel to the film surface, the emission shifts back to 580 nm and shows a strong polarization with a two-fold symmetry (Fig. 2f). The polarized emission maximum observed when the excitation is within the plane of the film (Fig. 2f), compared to its isotropy when perpendicular (Fig. 2e), is consistent with **2DPA-1** platelets oriented parallel to the substrate with a transition dipole along the long axis of the platelet. This molecular picture is further supported by grazing-incidence wide-angle X-ray scattering (GIWAXS), in which strong long-range orientational ordering but weak positional ordering is observed. A diffuse arc in q$_z$ axis represents an interlayer spacing in the *z* direction (Fig. 2g), and its peak around 1.69 Å$^{-1}$ in the 1D profile (Fig. 2h) corresponds to a spacing of 3.72 Å, which is close to our AFM observation (Fig. 1f). These results suggest that **2DPA-1**, when reconstituted as a spin-coated film, consists of discotic-shaped platelets aligned parallel to the substrate surface (Fig. 1b), which is also highly consistent with our AFM and TEM topology (Fig. S18 and S19).

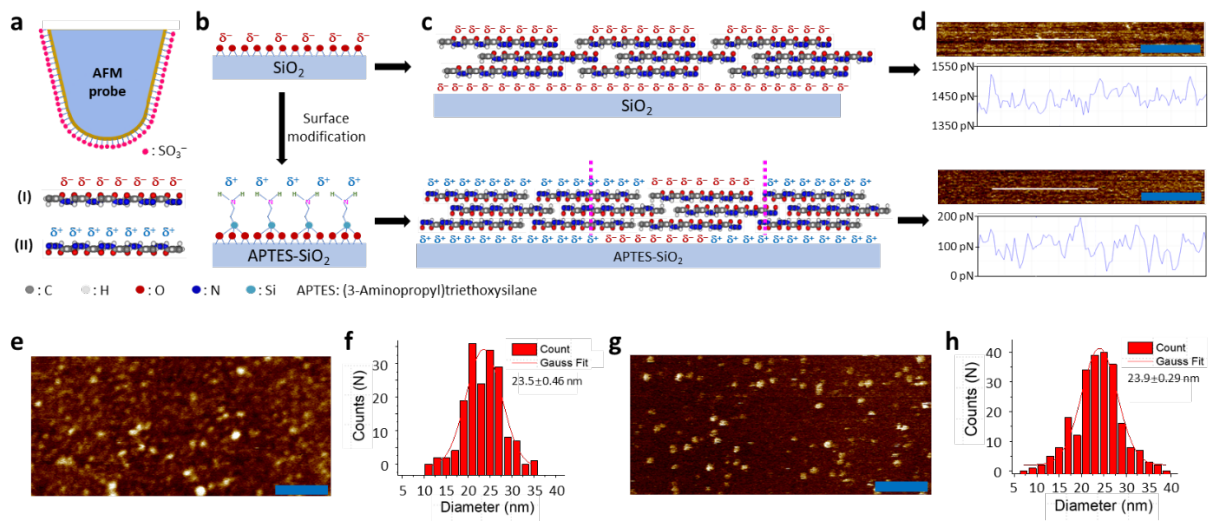

**Figure 3| Chemical force characterization of 2DPA-1 nanofilms. a,** A schematic representation of surface recognition by a sulfonate modified, negative charged AFM probe. In this molecular model, all amides are drawn vertically for better illustration. (I) Janus 2D molecule, negatively charged face; (II) Janus 2D molecule, positively charged face. **b,** Chemical modification of SiO$_2$ substrate flips the surface charge from negative (top) to positive (bottom). APTES: (3-aminopropyl)triethoxysilane. **c,** Surface recognition can be extended from the substrate surface to the film surface. Top: **2DPA-1** film on SiO$_2$; bottom: **2DPA-1** film on APTES-SiO$_2$. However, an imperfection modification could cause a local molecular flipping on the top surface (illustrated between the two pink dashed lines). **d,** Chemical force mapping of **2DPA-1** films on SiO$_2$ (top) and APTES-SiO$_2$ (bottom) substrates. Adhesion profiles along white lines are also given under images. Scale bar, 100 nm. **e,** Height image of a given area from a **2DPA-1** film on APTES-SiO$_2$ and its correlated adhesion force image (**g**). **e** and **g** were obtained simultaneously in one single scan under force mapping mode. Scale bar, 200 nm. **f** and **h,** Molecular size distributions from the height topology (**e**) and the adhesion force map (**g**). Red curve: Gauss fit of the distribution histogram.

To study the tacticity of amides inside of individual 2D platelets, we conducted a chemical Atomic Force Microscopy (cAFM) investigation of spin-coated nanofilms. If amides are oriented out-of-plane and isotactic, the **2DPA-1** should exhibit a net dipole orthogonal to the plane, leading to two distinct molecular surfaces (Fig. 3a, I and II). Alternatively, if amides are confined in-plane, syndiotactic, or atactic, the result is no detectable net dipole, giving two identical molecular surfaces (Fig. S35). These two scenarios can be distinguished by cAFM with a SO$_3^-$ terminated AFM probe (Fig. 3a) to detect the dipolar surface. We examined **2DPA-1** deposited on two entirely different substrates, one covered with electron-rich O atoms (SiO$_2$, Fig. 3b) and the other with electron-deficient protons (APTES modified SiO$_2$, Fig. 3b). The carbonyl of amides can either pair with the substrate or face the opposite direction, propagating this configuration throughout the aligned film such that the charged AFM probe can access the potential dipole (Fig. 3c). Indeed, cAFM shows two distinct adhesions for **2DPA-1** on either SiO$_2$ at 1307±64 pN or APTES-SiO$_2$ at 102±33 pN (Fig. 3d). This indicates a high degree of planar asymmetry for **2DPA-1** molecules. Interestingly, in the latter case, the height image (Fig. 3e) and force image (Fig. 3g) do not overlay, indicating a successful measurement of the adhesion force distinctly. The patches of high adhesive force indicate local molecular flipping due to imperfections in the APTES coating (between dash lines, Fig. 3c). The domain size estimated in this way (Fig. 3h) agrees with the molecular size distribution from height analysis (Fig. 3f) and single platelet diameters observed previously (Fig. 1f). This out-of-plane orientation, producing a dipole, indicates that **2DPA-1** should exhibit a high degree of

interlayer hydrogen bonding within the aligned platelets, a long-standing goal of nanomaterial synthesis.[32]

This interlayer hydrogen bonding for **2DPA-1** should translate into mechanical properties that exceed those of oriented platelet ensembles composed of 2D vdW materials such as graphene, hBN, and MoS$_2$ or comparable platelet dimensions.[20,33] We transferred homogenous and continuous nanofilms onto well-defined holey substrates (Fig. 4a) and measured their mechanical properties by AFM nanoindentation, a standard mechanical characterization method for 2D membranes.[3] Interestingly, we found that gas is captured underneath **2DPA-1** membranes during the wet transfer process and **2DPA-1** bubbles are formed from the sealed wells (Fig. 4a and 4b), similar to graphene systems studied by Bunch[34] and Sun[35]. Puncturing the bubble with an AFM tip at the periphery releases the trapped gas, yielding a super-flat, self-tensioned membrane (Fig. 4c). The pinholes stay localized during the indentation measurements (Fig. 4d), suggesting a degree of defect-tolerance for **2DPA-1**, not observed in graphene[3] and other conventional 2D materials.[36]

For a given elastic, free-standing 2D membrane, the indenting force ($F$) can be expressed as

$$F = \sigma_0^{2D}(\pi a)\left(\frac{\delta}{a}\right) + E^{2D}(q^3 a)\left(\frac{\delta}{a}\right)^3 \quad (1)$$

Where $\sigma_0^{2D}$ is the film pretension, $a$ is the diameter of the membrane, $\delta$ is the deflection at the center point, and $q$ is a dimensionless constant. This cubic stress-strain relationship is typical of 2D membrane systems such as graphene,[3] and exhibited by **2DPA-1** (Fig 4e). Here $E^{2D}$ is the Young's modulus of the material and can be obtained by fitting the force-displacement data to Eq 1 as shown in Fig 4f. We studied 120 pierced membranes from two samples with average film thicknesses of 12.8±1.4 nm and 33.9±2.2 nm using two diamond-like spherical AFM probes with two different tip radii (50±5 nm and 100±10 nm). Membranes are indented multiple times at the film center with different trigger forces (Fig. 4e). For all membranes, we observed no ultimate failure at the force curve limit but note several inconspicuous yield points and a continuous breaking process (Fig. 4e). Small plastic deformations shift subsequent force curves toward higher strain. Therefore, the intrinsic two-dimensional modulus is obtained from the elastic region corresponding to the first set of overlapping curves. The 2D yield strength ($\sigma^{2D}$) is further determined from $(FE^{2D}/4\pi R)^{1/2}$, where R is the tip radius and F is the force at the first yield point from which a force curve starts to lose its elasticity (Fig. 4e).

The measured elastic moduli of the membranes range from 30 to 90 GPa (Fig. 4f & 4g) with an average modulus of **2DPA-1** is 50.9±15.0 GPa, substantially higher than conventional unoriented thermoplastics (e.g. polycarbonate, 2.4 GPa), crosslinked polymers (e.g. toughened epoxy, 2.5 GPa), or oriented linear thermoplastics (e.g. nylon, 5 GPa), approaching that of metals (e.g. tin, 47 GPa and aluminum, 69 GPa) and high performance oriented linear polymers such as polyaramid (96 GPa) or ultrahigh molecular weight polyethylene (88 GPa).[37] As others have noted, an additional advantage of 2D polymers such as **2DPA-1** is that in the aligned form they exhibit isotropic stiffness within the 2D plane, doubling the effective stiffness when compared to 1D polymer counterparts that reinforce in only a single direction.[32] **2DPA-1** also exhibits an excellent yield strength of 0.976 GPa, with a standard deviation of 0.113 GPa. This yield strength is almost 4 times that of structural steel (ASTM A36, 0.25 GPa), in spite of **2DPA-1** having approximately 1/6 the density (Table S2).

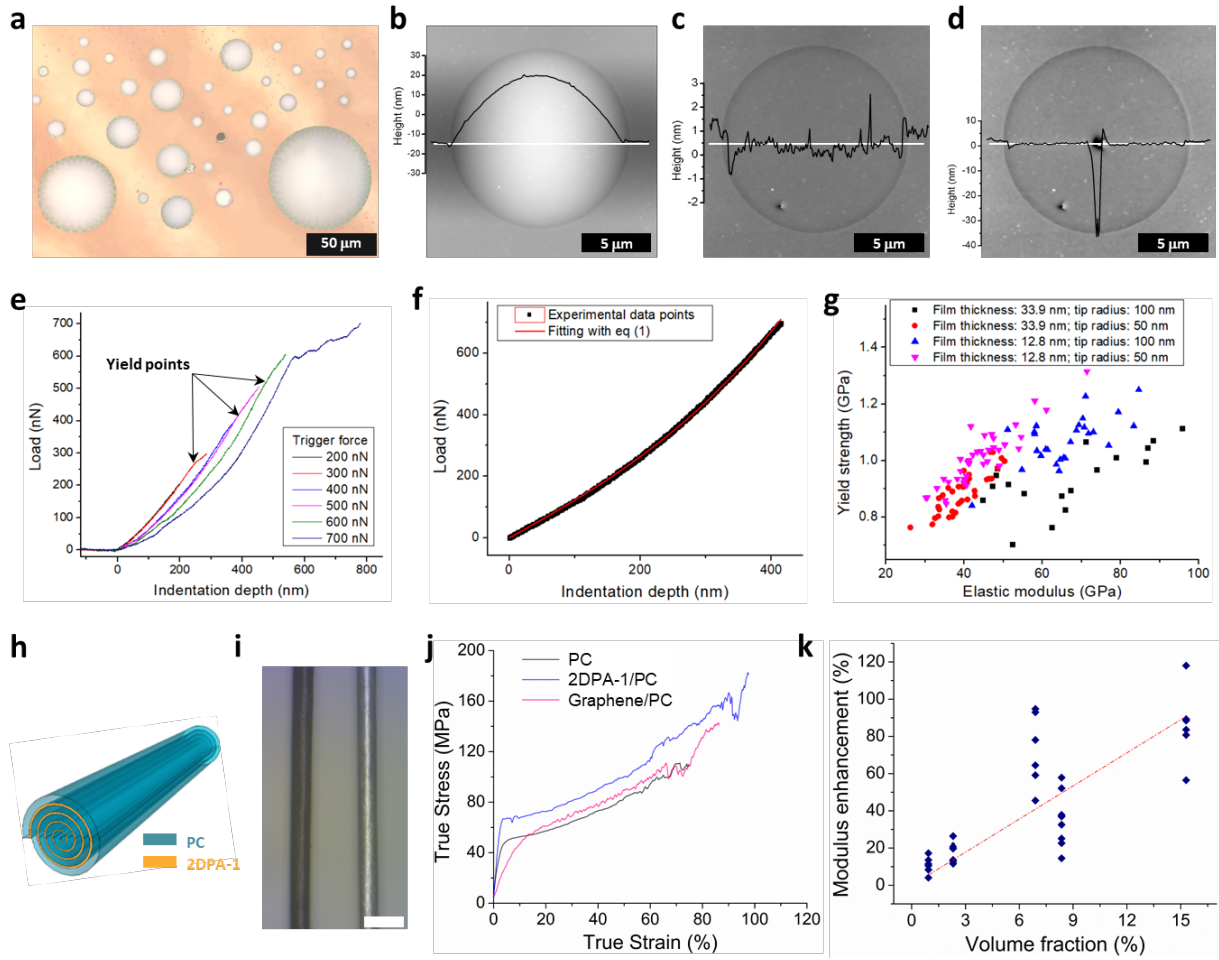

**Figure 4| Mechanical properties of 2DPA-1 nanofilms. a,** Optical micrograph of a 33.9-nm thick **2DPA-1** film on a Si holey substrate. White circles are intact membranes and the black dot in the middle represents a fractured membrane. **b,** A 13.7-μm SiO$_2$/Si well covered by a 12.8-nm thick **2DPA-1** film forming an impermeable bubble of trapped gas as shown by the AFM height profile (along the white line). **c,** Collapsed membrane after puncturing with an AFM tip and subsequent gas release, showing a flat height profile (along the white line). The small pinhole is located in the lower left. **d,** Membrane after indenting and its height profile along the white line. **e,** Indentation on a 12.8-nm thick, 13.2-μm wide free-standing membrane at its center with different trigger forces. Tip radius: 100 nm. Curve overlapping proves its elasticity and a right-shift of the curve indicates plastic deformation has occurred in the previous run. **f,** A representative force-displacement curve and its modulus fitting. Conditions: a 33.9-nm thick, 24.7-μm wide membrane with a 100-nm radius tip. **g,** Plot of 2D elastic modulus against its 2D yield strength of **2DPA-1**. **h,** Schematic illustration of an Archimedean scroll fiber. **i,** Optical micrograph of a hair (left) and a scrolled fiber (right). Scale bar, 100 μm. **j,** Representative true stress-strain curves from a 2D composite scrolled fiber, its polycarbonate (PC) control fiber, and a graphene/PC composite fiber (data reproduced from *Science* **2016**, *353*, 364). Volume fraction: **2DPA-1** /PC = 6.9%, Graphene/PC = 0.19%. **k,** Plot of modulus enhancement (($E-E_{PC}$)/$E_{PC}$) against different volume fractions of **2DPA-1**.

We further studied the mechanical response using conventional tensile testing methods. Normally those measurements are limited to macroscale and not applicable to nanomaterials. However, our previously established scroll fiber platform offers an opportunity to convert microscale mechanical properties into macro measurable quantities, and thus study the material behavior in real nanocomposite applications.[38,39] After layering an additional **2DPA-1** film onto a polycarbonate (PC) film and scrolling this nanostructure into an Archimedean nanostructured fiber (Fig. 4h and 4i), we found that the resulting fibers exhibit significant larger elastic moduli and tensile strength than PC controls, even at very low volume fraction ($V_{2DP}$). For instance, a 6.9% fraction of **2DPA-1** film enhances the fiber modulus by 72%, while the strength raises from 110 MPa to 185 MPa (Fig. 4j). Additionally, the **2DPA-1** composite fibers do not undergo "telescoping", an unscrolling phenomenon previously observed for graphene/PC scroll fibers with low initial modulus (Fig. 4j).[38,39] The integrity of the **2DPA-1** composites suggests a high-density of interfacial polar bonds available for adhesion to the PC matrix. In the absence of a telescoping mechanism, the composite fiber modulus is a linear combination of PC matrix and 2D polymer, written as Eq 2.

$$E = (1 - V_{2DP})E_{PC} + V_{2DP}E_{2DP} \qquad (2)$$

$$\eta = \frac{E - E_{PC}}{E_{PC}} = (\frac{E_{2DP}}{E_{PC}} - 1)V_{2DP} \qquad (3)$$

Plotting the modulus enhancement as a function of **2DPA-1** volume fraction should result in a linear correlation (Fig. 4k), and the slope can be used to estimate the modulus ratio $\eta = E_{2DP}/E_{PC} = 6.1$ (Eq 3). Although approximate, the high apparent modulus of the **2DPA-1** phase, combined with its propensity to adhere to the matrix, suggest that 2DP-based nanocomposites could provide a significant advance in mechanical performance compared to legacy nanocomposites reinforced with inorganic van der Waals nanoplatelets.

In conclusion, we have discovered an irreversible, solution-phase polymerization that promises new families of mechanically and chemically stable 2D polymers, analogous in properties to their 1D organic counterparts. Unlike reversible synthetic routes, the approach in this work is amenable to organic chemistry rather than a crystallization process. We find, as predicted[32] that the polyaramid system introduced in this work has extra-ordinary mechanical properties, almost 4 times stronger than structural steel, exhibiting great potential for composite materials as well as lightweight coatings. We envisage that the 2D polyaramid system we describe herein could be further structurally tuned, paving the way for a new generation of polymer materials as barrier coatings, lightweight structure reinforcement, nanofiltration, and gas separation.

**Methods**

**Synthesis of 2DPA-1.** A 40 mL vial equipped with a stir bar was added with 126 mg of melamine (1 mmol, 1 equiv.), $CaCl_2$ (0.5 g), and 265 mg of trimesic acid trichloride (1 mmol, 1 equiv.), followed by 9 mL of N-Methyl-2-pyrrolidone and 1 mL of pyridine. The mixture was stirred at room temperature. After 16 hours, the whole reaction mixture became a gel. This gel was cut into small pieces and then soaked in EtOH (80 mL), followed by 30 min bath sonication (if necessary). The resulting cloudy mixture was further filtered or centrifuged, followed by deionized $H_2O$ (80 mL) and acetone (80 mL) washing. A pale-yellow solid (228 mg, 81%) was received after house-vacuum drying at 100°C for 16h.

**Preparation of 2DPA-1 nanofilms. 2DPA-1** powder was dissolved in trifluoroacetic acid (TFA), forming a homogenous solution. To a clean $SiO_2$-covered (300 nm) Si wafer, **2DPA-1** solution was added on top.

Then this wafer was spun at a certain rate for 1 min, giving a uniform nanofilm. Its thickness can be measured by AFM at scratches made by a fine needle (Supplementary Fig. S11).

**Atomic force microscopy.** AFM imaging was performed on *Asylum Research* instruments (Cypher S and MFP-3D) and a *Bruker* Veeco Multimode 8 instrument in AC mode using various probes (Arrow UHF, NPG-10, AC-160, and FASTSCAN-D-SS) for different tasks. Data was processed using the *Gwyddion* software package and built-in software in Asylum and Bruker systems.

Chemical force mapping was performed on a *Bruker* Veeco Multimode 8 instrument. To eliminate the influence of the surface water layer and contaminations, all peak force measurements were done under fluid mode using a liquid cell. Deionized water was used as an experimental medium. Gold coated AFM probes were modified by sodium 3-mercapto-1-propanesulfonate in EtOH. To eliminate the influence from different probes, experimental results, including substrate controls, were obtained in one measurement without changing probes.

Nanoindentation was performed on a Cypher S instrument. Si holey substrates were fabricated by photolithography and diamond-like spheric probes (*Biosphere* NT_B50_v0010 and NT_B100_v0010) were purchased from *Nanotool AFM Probes*. Film thicknesses were obtained by taking an average at 7 different positions nearby the place of interest. After manual calibration, the membrane of interest was imaged under tapping mode and punctured at its periphery under contact mode. Then the tip was moved to the film center and nanoindentations were performed with different trigger forces. The loading force-displacement curves were extracted and analyzed, offering intrinsic 2D modulus and yield strength of **2DPA-1**.

**Polarized photoluminescence measurement.** The whole optical setup is shown in Supplementary Fig. S20. A continuous-wave 532 nm laser (Edmund, 35-072) was used for excitation. The incident light travelled through a linear polarizer and a half-wave plate (mounted on a motorized stage) and focused onto the sample using an objective lens (Zeiss, 100x, NA=0.75). The angle of the half-wave plate was adjusted to maximize photoluminescence intensity. Then, the stage was moved a few μm away because **2DPA-1** has already photobleached to some extent during the focusing and adjustment of the half-wave plate. The signal was collected with a spectrometer (Princeton Instruments, Acton SpectraPro SP-2150, and PyLon). The excitation power for photoluminescence measurements was 500 μW and the exposure time was 10 seconds. For simple PL measurement, a spectrometer was used for data collection. However, for the polarized PL study, we used an EMCCD camera (Andor, iXon3), which is much more sensitive, to trace a longer time course despite the photobleaching of **2DPA-1**. The polarity of the incident light was controlled by rotating the half-wave plate and the PL signal was collected every 5 degrees with 5 seconds exposure time. The excitation power was 2 μW for excitation polarization. All measurements were conducted at room temperature under air.

**Scrolled fiber test.** The tensile test was performed on an Instron 8848 Micro Tester. Firstly, the scrolled fiber was glued onto a hollow cardboard using epoxy resin, with a gauge length of 16mm. Then the whole sample was mounted onto the micro tester, and the connecting parts on the cardboard were cut, leaving a free-standing scroll fiber. The test was carried out at room temperature with a strain rate of 0.1 mm/s using a 10-N load cell. The force-displacement curve was recorded until the fiber breaks off (Supplementary Fig. S40).

**Data availability**

The data that support the findings of this study are available from the corresponding author upon reasonable request.

**Acknowledgments**

This work was funded by the Army Research Laboratory under cooperative agreement W911NF-18-2-0055. HJK holds a Career Award at the Scientific Interface from the Burroughs Wellcome Fund, which supported this work. We acknowledge the Center for Nanoscale Systems at Harvard, a member of the National Nanotechnology Coordinated Infrastructure Network (NNCI), which is supported by the National Science Foundation under NSF award no. 1541959. This research used beamline 11-BM Complex Materials Scattering (CMS) of the National Synchrotron Light Source II (NSLS-II) and the Center for Functional Nanomaterials (CFN), both of which are U.S. Department of Energy (DOE) Office of Science User Facilities operated for the DOE Office of Science by Brookhaven National Laboratory under Contract No. DE-SC0012704. We thank Dr. Esther Tsai for her assistance in performing experiments at the beamline and Prof. Rafael Verduzco for the help on beamline access.

**Author contributions**

Y.Z. conceived and designed the reaction system, synthesized, and characterized **2DPA-1**. P.G., Y.Z., and J.T. performed HR-AFM measurement and data analysis. Y.Z. designed and T.I. conducted polarized PL measurement. Y.Z. designed and P.G. conducted chemical force microscopy measurement. Y.Z. measured mechanical properties by nanoindentation at MIT and S.-R.E. confirmed at ARL. Y.Z. performed the scrolled fiber test. T.I. and X.G. helped with the data analysis. G.Z., D.K., J.Y., Z.Y., and H.J.K. helped with the theory and simulation. M.K., P.L., A.T.L., and J.Y. offered substrates. Y.Z. and M.S.S. co-wrote the manuscript. All authors contributed to discussions.

**Competing interests**

The authors declare no competing interests.

**Additional information**

Correspondence and requests for materials should be addressed to M.S.S.

**Reprints and permissions information** is available at http://www.nature.com/reprints.

Supplementary Materials for

# An Irreversible Synthetic Route to an Ultra-Strong Two-Dimensional Polymer

Yuwen Zeng[1], Pavlo Gordiichuk[1], Takeo Ichihara[1], Ge Zhang[1], Sandoz-Rosado Emil[2], Eric D. Wetzel[2], Jason Tresback[3], Jing Yang[1], Zhongyue Yang[1], Daichi Kozawa[1], Matthias Kuehne[1], Pingwei Liu[1], Albert Tianxiang Liu[1], Jingfan Yang[1], Heather J. Kulik[1], Michael S. Strano[1]*

Department of Chemical Engineering, Massachusetts Institute of Technology, Cambridge, MA 02139, USA

U.S. Army Research Laboratory, Aberdeen Proving Ground, MD 21005-5069, USA

Center for Nanoscale Systems, Harvard University, Cambridge, MA 02139, USA

# Table of Contents

# Chemical Kinetic Modeling of 2D vs 3D Solution Polymerization

Homogeneous polymerization strongly disfavors the generation of 2D polymers, due to the unfavorable scaling of growth rate compared to 3D random polymers, and the lack of error correction.[10]

To elaborate on the scaling of growth rate, we may abstract this problem to consider monomers as structure-less beads of radius $r_m$ and volume $V_m$. As the growth of the planar molecule comprised of i monomers (denoted as $G_i$) proceeds, the number of reactive sites along its periphery increases and hence so does the rate constant for addition. This is a fundamental difference compared to linear polymers. To determine this scaling, consider the radius and area of $G_i$ as $r_i$ and $A_i$ respectively. The radius and area of the corresponding monomers are $r_m$ and $A_m$ respectively. Naturally,

$$r_i = \sqrt{\frac{A_i}{\pi}} \sim \sqrt{\frac{iA_m}{\pi}}$$

while the circumference is $2\pi r_i$. For large i, we can approximate the arc length occupied by a monomer as $2r_m$. Hence, the number of peripheral sites $S_i$ available for addition on $G_i$ grows as the square root of i:

$$S_i \approx \frac{2\pi r_i}{2r_m} = \frac{2\pi\sqrt{\frac{iA_m}{\pi}}}{2\sqrt{\frac{A_m}{\pi}}} = \pi\sqrt{i}$$

In terms of the rate constant for monomer addition per site, $k_o$, we expect the constant for the i+1 addition to be

$$k_i \approx k_o \pi \sqrt{i}$$

Meanwhile, the addition of a monomer to a 2D polymer $G_{i-1}$ can also result in a defective 3D structure, which we label as $D_i$. The 3D random growth imparts a different rate constant scaling as we consider the volume and projected area $V_m$ and $A_m$ of the monomer, with corresponding $V_{Di}$ and $A_{Di}$ for the volume and surface area of $D_i$.

$$r_{Di} = \left(\frac{3V_{Di}}{4\pi}\right)^{\frac{1}{3}} \approx \left(\frac{3iV_m}{4\pi}\right)^{\frac{1}{3}}$$

Hence, the number of reaction sites $S_{Di}$ on the surface of the growing sphere $D_i$ scales as

$$S_{Di} \approx \frac{A_{Di}}{A_m} = \frac{4\pi r_{Di}^2}{\pi r_m^2} = \pi (i)^{\frac{2}{3}}$$

We then expect the rate constant for defective 3D random growth to scale as

$$k_{Di} \approx k_o \pi (i)^{\frac{2}{3}}$$

Given the discussion above, when all things being equal, the defect structure will grow faster with i to the 1/6 power (orange and blue solid straight lines, **Figure S1a**).

The geometric analysis of the number of reaction sites is supported by kinetic Monte Carlo (kMC) simulation results. As shown in **Figure S1a** (orange and blue dotted lines), the number of reactions sites for a 3D random polymer scales with $i^{0.73}$, which is faster than polymer confined to a 2D plane by $i^{0.16}$, consistent with the conclusion in the main text. The slight deviation from the ideal value is due to voids in the polymer that reduces its size, and the winding circumference leading to an increased perimeter length. **Figure S1b** shows that the addition of a monomer from a direction out of the polymer plane can prevent the planar ring closure reaction, thus leading to a permanent structure change from 2D to 3D.

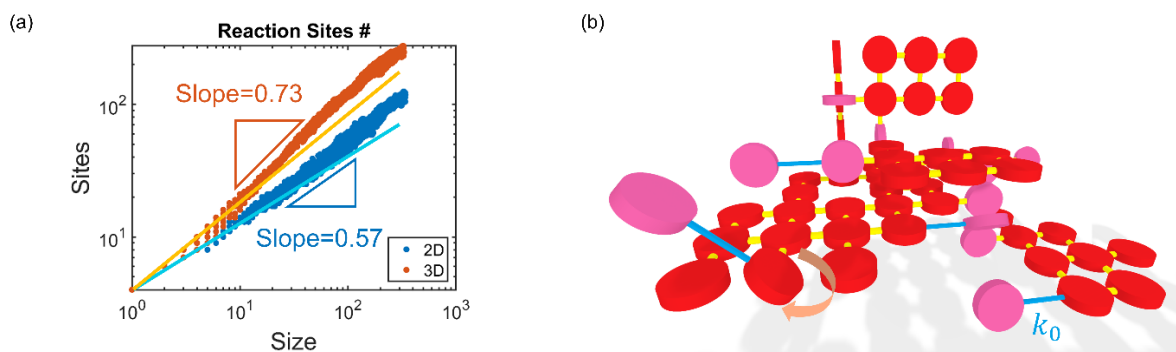

**Figure S1.** Schematics and scaling law. **a**, Reaction sites scaling with polymer size for 2D versus 3D polymer, blue solid lines are results from the theoretical scaling laws and dots are from kMC simulations. **b**, Schematics of 2D versus 3D growth, with out-of-plane bond formation illustrated by blue sticks, defective units are shown in pink rather than red.

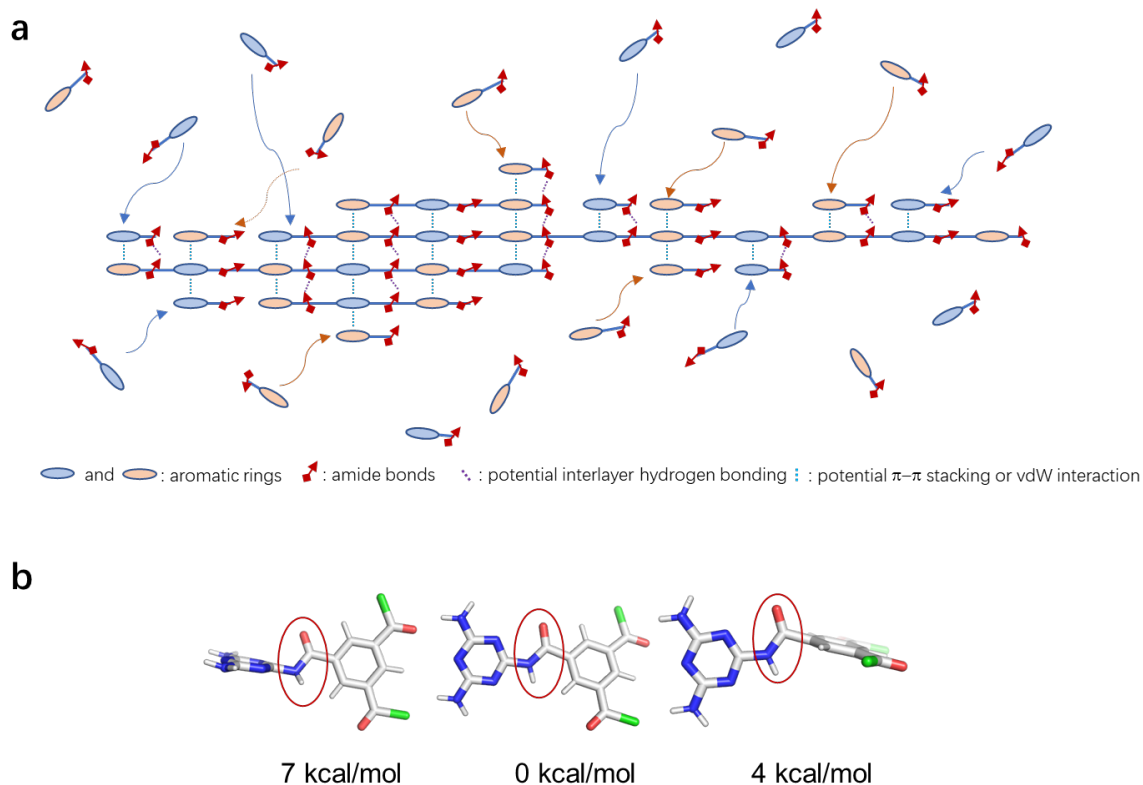

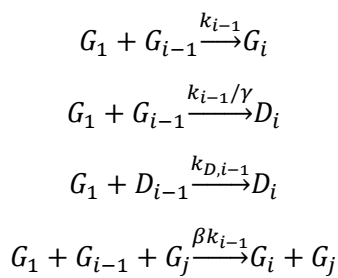

7 kcal/mol    0 kcal/mol    4 kcal/mol

**Figure S2.** Schematic illustration of auto-catalysis by self-templating (**a**) and linkage in-plane tendency arising from linkage-core conjugations (**b**). Computational method used in **b**: Gas-phase geometry optimizations with Q-Chem v4.2 to compute 298 K conjugation enthalpies employed the ωB97-XD/6-311+G(d,p) DFT functional and basis set combination.

However, certain mechanisms could effectively suppress the defect formation and/or 3D growth. We have simulated two of these mechanisms, namely the autocatalysis and bond-planarity effects, characterized by two dimensionless parameters **β** and **γ** respectively (**Figure S2**). Where **β** represents a rate constant ratio of auto-catalyzed 2D growth pathway (**Figure S2a**) and non-catalyzed 2D growth pathway, and **γ** stands for an in-plane tendency of given linkages (**Figure S2b**), which equals to a probability ratio of in-plane and out-of-plane states. The reaction networks are shown below:

$$G_1 + G_{i-1} \xrightarrow{k_{i-1}} G_i$$

$$G_1 + G_{i-1} \xrightarrow{k_{i-1}/\gamma} D_i$$

$$G_1 + D_{i-1} \xrightarrow{k_{D,i-1}} D_i$$

$$G_1 + G_{i-1} + G_j \xrightarrow{\beta k_{i-1}} G_i + G_j$$

Here, $G_1$ is the monomer, $G_i$ is the 2D polymer with i monomer units, $D_i$ is the 3D polymer with i monomer units. The irreversible rate of addition is $k_i$ and $k_{D,i}$ for 2D and 3D polymers respectively, where the former increases with $i^{1/2}$, and the latter increases with $i^{2/3}$. Parameter **γ** controls the rate of transformation from 2D to 3D defective polymer, and **β** dictates the rate ratio of auto-templated 2D growth and normal 2D growth.

In **Figure S3**, we show that both the average size and yield of 2D polymers are significantly enhanced as **β** and/or **γ** increases. The yield is above 90% even for moderate values of **β**=100 and **γ**=1000. Our calculations show for the first time from theory, the feasibility of producing two-dimensional polymers from irreversible polymerization in solution.

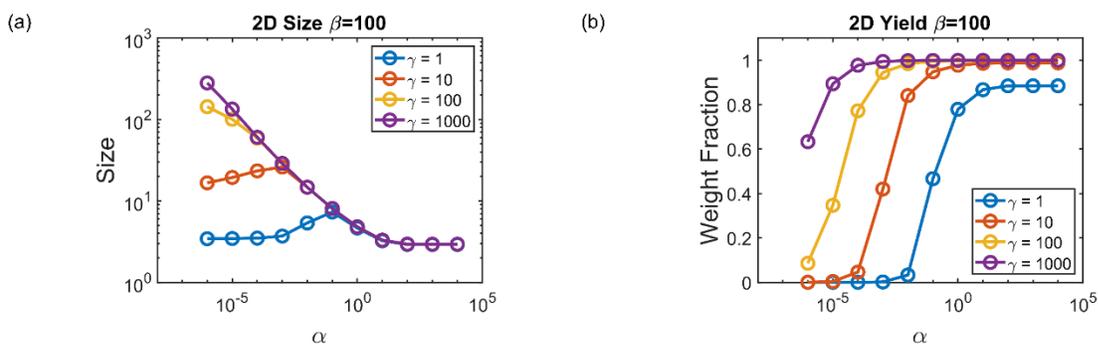

**Figure S3.** A combined effect of bond-planarity and autocatalysis. **a**, Average size of 2D polymers at different values of **γ** with **β** = 100. **b**, Yield of 2D polymers at different values of γ with **β** = 100. Parameter **α** describes the rate of monomer activation, which is not discussed here.

# Synthesis and Characterization of 2DPA-1 and 2DPA-Am

**Materials**

Chemical reagents (melamine, trimesoyl chloride, isothaloyl chloride, $CaCl_2$, and pyridine) and anhydrous solvents (N-methyl-2-pyrrolidone, acetone, and trifluoroacetic acid) were purchased from *Aldrich* and used as received. For convenience, syntheses were conducted using standard Schlenk techniques or in an inert atmosphere glovebox unless otherwise stated. However, all starting materials can be also weighted and mixed in an ambient atmosphere and then sealed with a cape.

Thermal oxide wafers ($SiO_2$/Si, oxide thickness: 300 nm) were purchased from *Waferpro* and diced into certain sizes. TEM grids, highest grade V1 Mica discs, ultra-flat Si and $SiO_2$ substrates were obtained from *Ted Pella*. AFM probes (Arrow UHF, NPG-10, AC-160, and FASTSCAN-D-SS) were purchased from *Oxford Instruments*, *Bruker*, *Olympus*, and *NanoWorld*.

Polycarbonate (PC, granule) used in this research was purchase from *Aldrich* with an average molecular weight of 60K.

**Characterization techniques**

Thermogravimetric analysis (TGA) was operated on a *Discovery TGA-1* instrument under $N_2$ flow. Fourier-transform infrared (FTIR) measurements were performed by using a *Bruker* ATR-FTIR Spectrometer with a reflection diamond ATR module. Powder X-ray diffraction (PXRD) data was recorded on a *PANalytical X'Pert Pro* diffractometer using a Cu target (Kα1 radiation, λ = 1.54059 Å). Atomic force microscopy (AFM) images were collected using *Asylum MFP-3D*, *Asylum Cypher S*, and *Bruker Veeco Multimode 8* instruments and analyzed with *Gwyddion* or *Cypher*. Scanning electron microscopy (SEM) images were collected on a *Helios 660* from *FEI* and a *Sigma 300 VP* from *Zeiss*. Wide-angle X-ray scattering (WAXS) patterns were acquired on beamline 11-BM Complex Materials Scattering (CMS) of National Synchrotron Light Source II (NSLS-II) in the Brookhaven National Laboratory. $N_2$ sorption measurements were carried out on a Micromeritics ASAP 2020 System at 77K using a liquid $N_2$ bath.

**Synthesis and Purification of 2DPA-1 (*also denoted as YZ-2 in the whole SI*)**

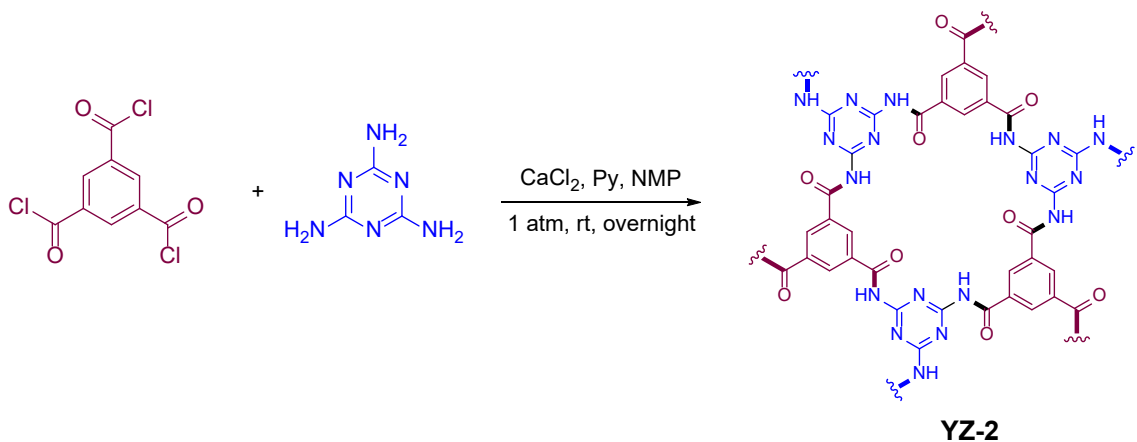

**Figure S4**. Synthetic scheme of **2DPA-1** (also denoted as **YZ-2** below).

To a 40 ml glass vial equipped with a stir bar, trimesoyl chloride (265 mg, 1 mmol, 1 equiv) and melamine (126 mg, 1 mmol, 1 equiv) were added followed by CaCl$_2$ (500 mg), anhydrous NMP (9 mL), and pyridine (1 mL). The reaction mixture was vigorously stirred overnight at room temperature. During the reaction course, the whole reaction system became a gel. This gel was cut into small pieces, mixed with 80 mL of ethanol, and stirred/sonicated to give a cloudy mixture. The resulting mixture was further filtrated or centrifuged, followed by H$_2$O (80 mL) and acetone (80 mL) washing. A pale-yellow solid was obtained after house-vacuum drying at 100°C for 8h.

**Note**: Although most of our synthesis was operated under N$_2$ atmosphere in a glovebox, it is just for convenience (most chemicals and anhydrous solvents were stored in glovebox), not mandatory. Starting materials are stable enough to weigh in air and the 2D condensation is not sensitive to O$_2$. We observed no difference when the reaction is carried out in air and sealed with a cap.

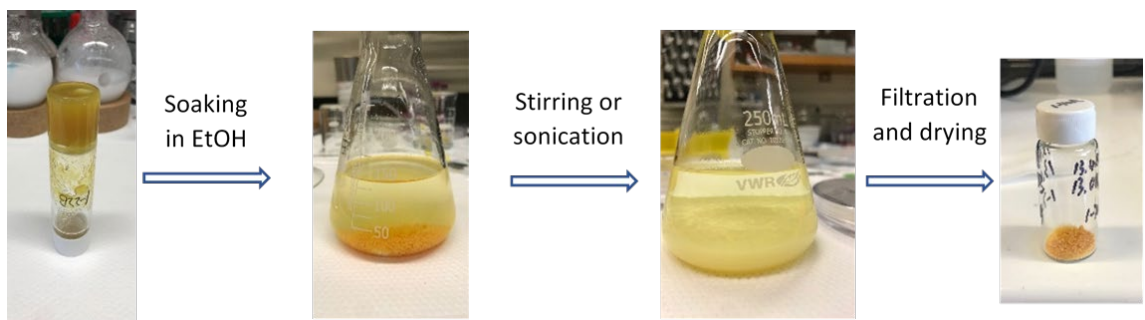

**Figure S5**. Reaction work-up and purification of **YZ-2**

**2DPA-Amorphous** (also denoted as **YZ-Amorphous**) is synthesized when trimesoyl chloride (1 equiv) is replaced by isothaloyl chloride (1.5 equiv) under standard conditions and the same purification method was used (**Figure S6**). **YZ-Amorphous** serves as an amorphous control in this study.

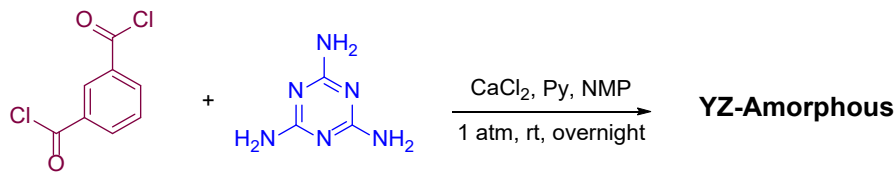

**Figure S6**. Synthesis of **2DPA-Amorphous**

**Chemical stability and solubility of YZ-2 in strong acid**

It is well known that most 2D materials made from reversible chemistries (2D COFs for example) possess low chemical stabilities, especially towards strong acids.[10,40] However, in organic chemistry, amide bonds are considered as stable chemical bonds under ambient conditions[41] (rate constant for hydrolysis are of the order of $10^{-11}$ s$^{-1}$), as well as in acidic and basic aqueous solutions.[42] Moreover, polyaramid such as *p*-phenylene terephthalamide (PPTA, also known as Kevlar) can stand in 100% of sulfuric acid at 80°C for at least 30 min during the wet-spinning process.[43]

During our investigation, we made several **YZ-2** stock solutions (2-5 mg/mL in trifluoroacetic acid, TFA). No oxygen or water-free operation is needed though caps with PTFE lining are used to prevent the solvent vapor from escaping. We observed no significant degradation over a long period (> 3 months) during our HR-AFM and spin-coated film studies.

**YZ-2** is insoluble in water and common organic solvents, however, it dissolves quickly in TFA to form a clear, pale yellow solution. No residue is observed in this solution. Dark-field dynamic light scattering study[44] was performed and its result indicated that there are no particles larger than 30 nm (detect limit) in a 1 mg/mL solution. Given that the molecular size of **YZ-2** is around 10 nm, we conclude that there is no molecular aggregations but only single molecules in the solution.

**Mechanical stability of YZ-2**

**YZ-2** is stable to bath sonication, which is widely used for reaction work up, purification, and dispersion. No obvious adverse effects were observed.

**Thermogravimetric Analysis**

**Method**: Few milligrams of samples were placed in a high-temperature Pt pan and mounted on a *Discovery TGA-1* instrument. The measurement was done under $N_2$ flow with a ramp rate of 5 degrees per minute.

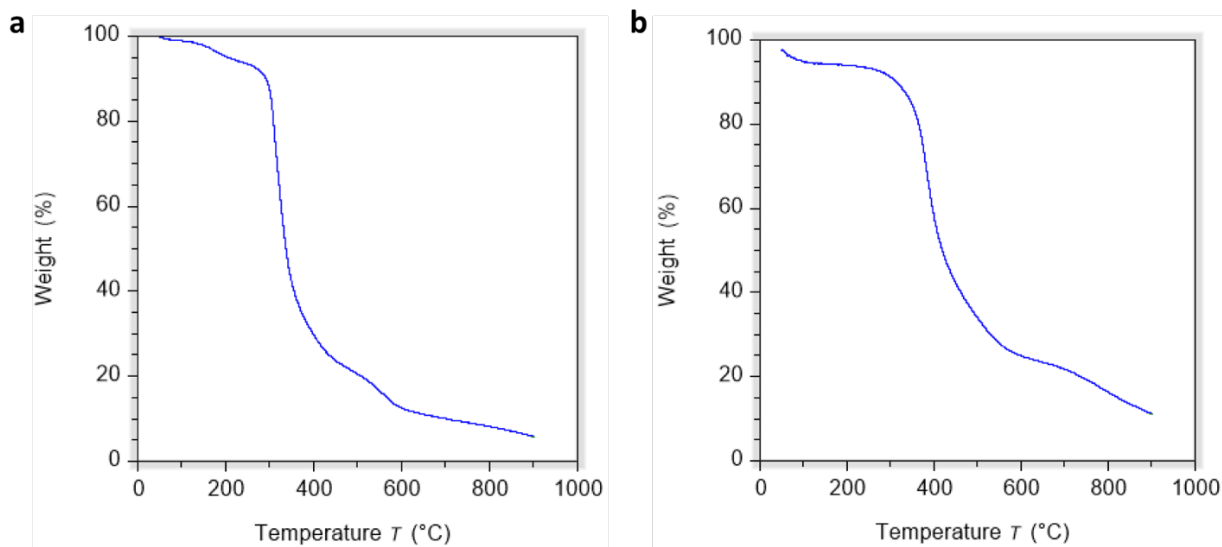

**Figure S7**. Thermogravimetric analysis of **YZ-Amorphous** (**a**) and **YZ-2** (**b**). **YZ-Amorphous** is synthesized when trimesoyl chloride is replaced by isothaloyl chloride under standard conditions. Samples were measured under $N_2$ flow. Ramp rate: 5°C /min to 900°C; isothermal at 50°C for 1 min.

A comparison between melamine (starting material), **YZ-Amorphous**, and **YZ-2** TGA curves and their first derivatives (DTG curves) were also shown in **Figure S8**. Trimesoyl chloride (TC) is not measured and shown here because even there is a small amount of TC left in the reaction mixture, it would be dissolved and thus washed away in the following purification processes.

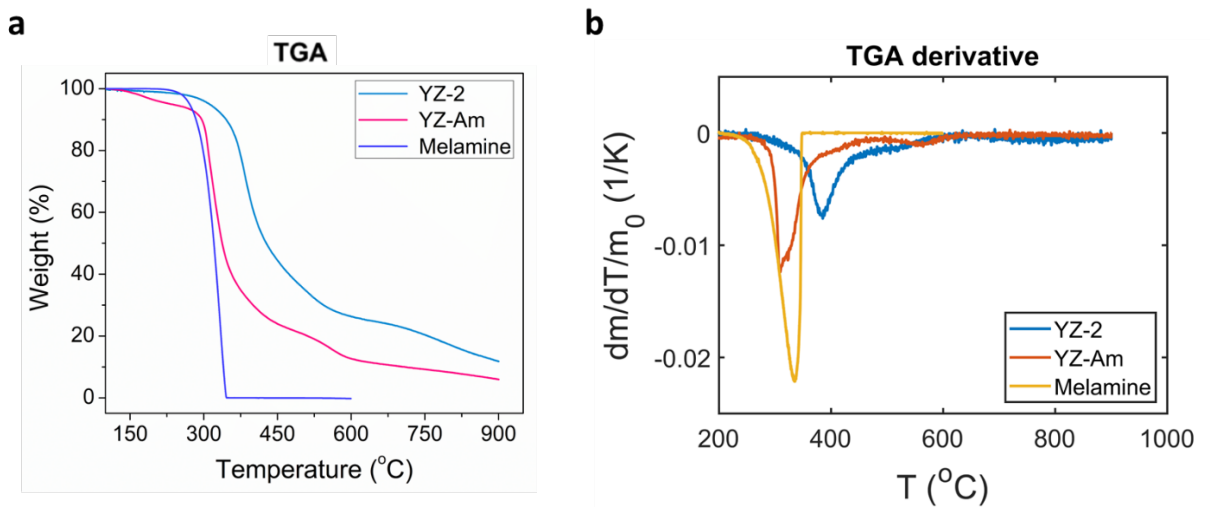

**Figure S8**. Thermogravimetric analysis (TGA) curves of melamine, **YZ-Amorphous** and **YZ-2** (**a**) and their derivative thermogravimetry (DTG) curves (**b**). Samples were measured under $N_2$ flow. Ramp rate: 5°C /min to 900°C; isothermal at 50°C for 1 min.

**Purity estimation:** The purity (weight fraction) of crystalline 2D polymer in the reaction product could also be estimated from the thermogravimetric analysis (TGA) data. We assume that the **YZ-2** sample contains both crystalline 2D polymer and amorphous 3D polymer, where the 2D one is the desirable "pure" product. We can then approximate the TGA behavior of amorphous 3D polymer with that of the sample **YZ-Am**. The TGA curve of a pure crystalline 2D polymer can be calculated by subtracting the contribution of amorphous 3D polymer from the whole TGA curve of **YZ-2**, providing we already know the purity. An upper bound of amorphous weight fraction (denoted as x) can be drawn, as the weight change ($\frac{dm}{dT}$) should be negative for both types of 2D polymers. Therefore, we can derive the following equation:

$$(1-x)\frac{d}{dT}m_{\text{YZ-Cry}} = \frac{d}{dT}m_{\text{YZ-2}} - x\frac{d}{dT}m_{\text{YZ-Am}} < 0$$

The differential weight change curves of crystalline 2D polymer, with different assumptions of amorphous weight fractions (x) are shown in **Figure S9**. The weight fraction of amorphous impurity should be bounded by 5%, otherwise the weight change ($\frac{dm}{dT}$) would be positive as depicted below.

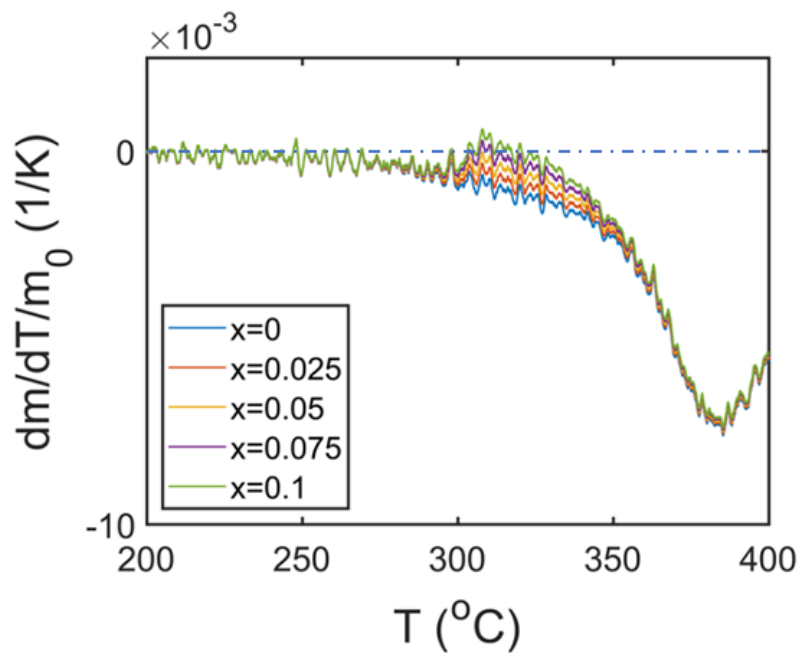

**Figure S9.** Differential thermogravimetric curves of pure crystalline 2D polymer derived using the above equation with different assumptions of amorphous weight fraction x.

**Thermal stability of YZ-2 nanofilms:** We found that suspended nanofilms are stable in 300°C for at least 1min without any obvious damage be observed.

**Fourier-Transform Infrared (FT-IR) Spectroscopy**

All FT-IR spectra were collected on a *Bruker* ATR-FTIR Spectrometer with a reflection diamond ATR module. The FT-IR spectrum of **YZ-2** clearly indicates that all starting materials were consumed or removed after reactions.

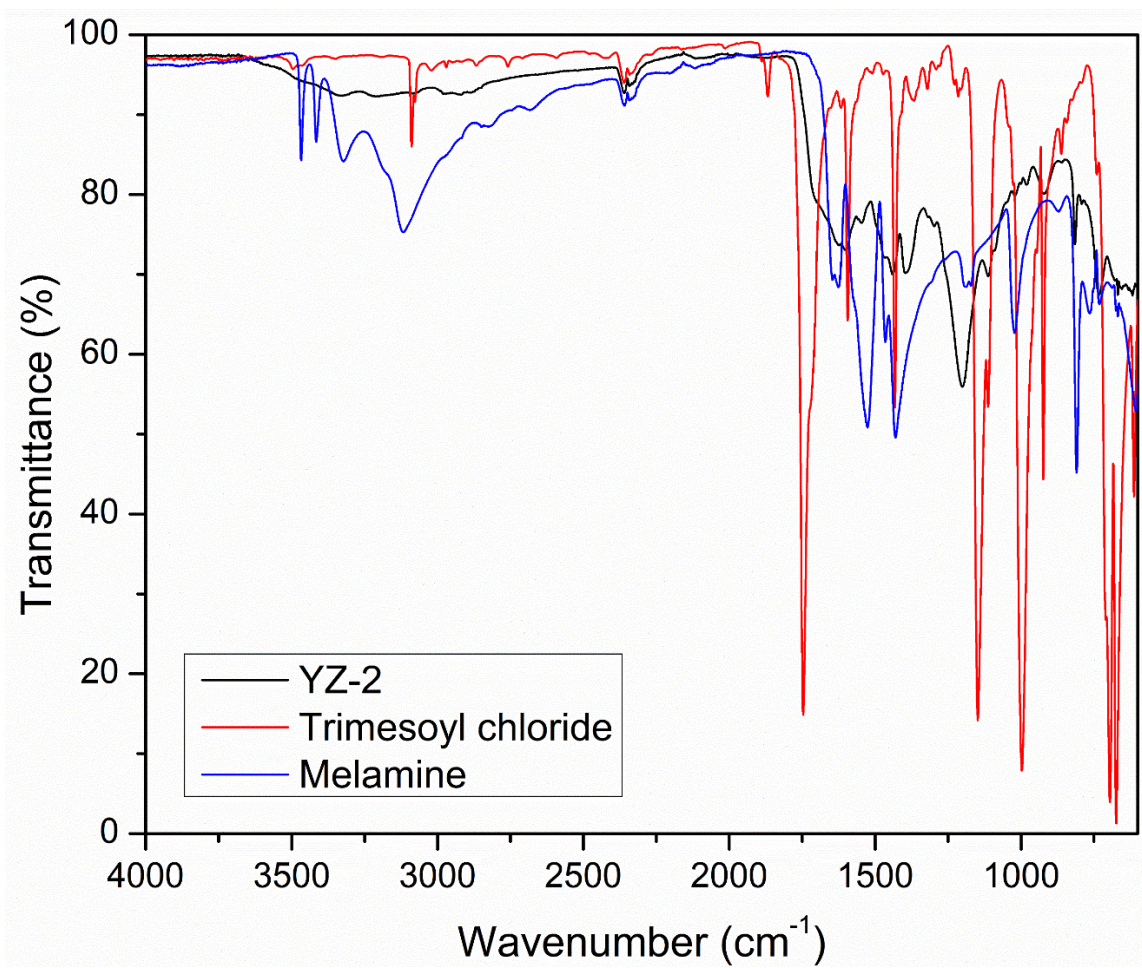

**Figure S10**. FT-IR spectra of **YZ-2** (black), trimesoyl chloride (red), and Melamine (blue)

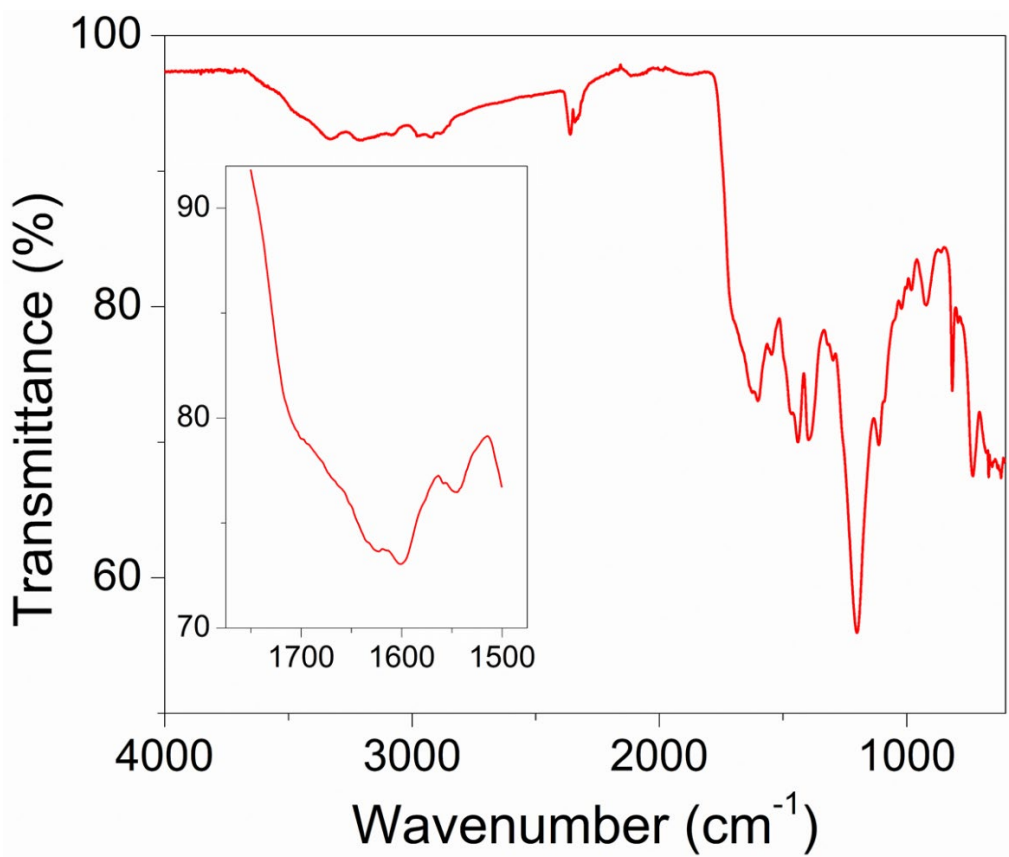

**Figure S11**. The whole FT-IR spectrum of **YZ-2** and its amide region is shown in the inset.

**Powder X-ray Diffraction (PXRD)**

**Characterization detail: YZ-2** powder was grounded and placed onto a spinning zero-background Si substrate. PXRD measurement was then performed on a *PANalytical X'Pert Pro* instrument using a Cu target (Kα1 radiation, λ = 1.54059 Å). Sample stage: Open Eularian Cradle (OEC); Temperature: 25°C; 2 Theta range: 5-60 degree.

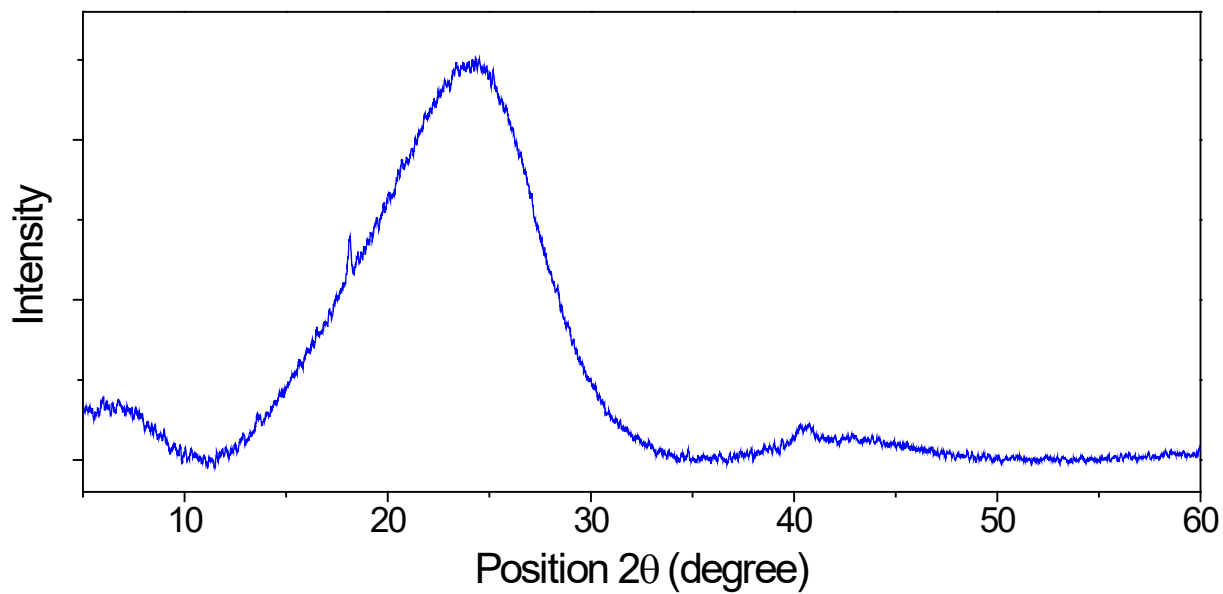

**Figure S12**. Powder X-ray diffraction (PXRD) of **YZ-2** powder

The peak centered at 24.3 degrees, according to Bragg's law, it corresponds to an average interlayer spacing of 3.66 Å.

**High-Resolution Atomic Force Microscopy (AFM) Characterization**

**Method**: Images were collected using a *Cypher S* (Asylum Research Oxford Instruments) AFM, or a *Veeco Multimode 8* (Bruker) AFM in AC mode. Data was analyzed with *Gwyddion* or *Asylum* AR*14* (Cypher's built-in analysis module). Ultra-high frequency proves (*Arrow UHF*) or Bruker FASTSCAN-D-ss probes were used under blue drive mode with a small laser.

**Controlled experiment**: Mica was measured directly (**Figure S13**) and showed an ultra-flat surface with an RMS roughness less than 40 pm over a 5*5 um area (normally 20-40 pm, calculated from *Asylum AR14*).

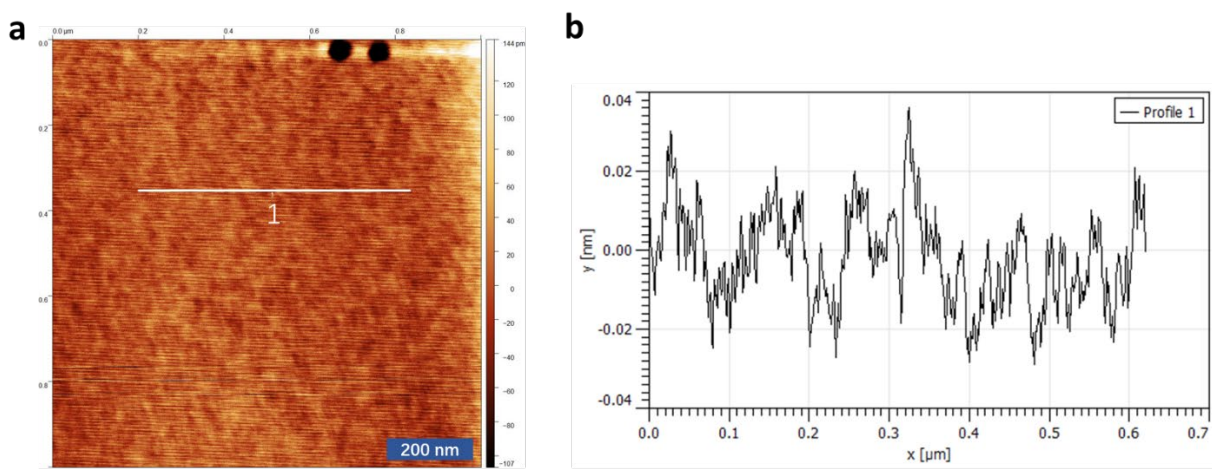

**Figure S13**. AFM topologic image of a freshly cleaved mica surface (**a**) and its height profile along the white line (**b**).

**Sample preparation:** Mica is freshly cleaved and used immediately as an ultra-flat substrate. **YZ-2** powder was dissolved in TFA (0.01 mg/mL) and spin-coated (2000 rpm) onto a mica substrate. The sample was then immersed in water several times and scanned with an *Asylum Cypher S* AFM in AC mode.

In most cases, due to the strong intermolecular interaction, even at very low concentrations, single platelets tend to sit on another single molecule rather than the bare mica surface, leading to the formation of bilayer clusters. Indeed, bilayers are observed predominately (**Figure S16**) while single molecules are rare and hard to find (**Figure S14 & 15**). By increasing the solution concentration (0.1 mg/mL or higher), discontinuous bilayer nanoclusters, and thin films start to show up (**Figure S17**).

**Note**: Samples are prepared in situ and examined immediately. Contaminations (presumably hydrocarbons) and sample changes (creeping of molecules) are observed even after few hours.

**Data analysis**: One single-molecule image is shown in **Figure S14**. Its height is measured at three different places. The mean values of the mica and platelet surfaces are determined by *Gwyddion*, using zero-order polynomial fitting.

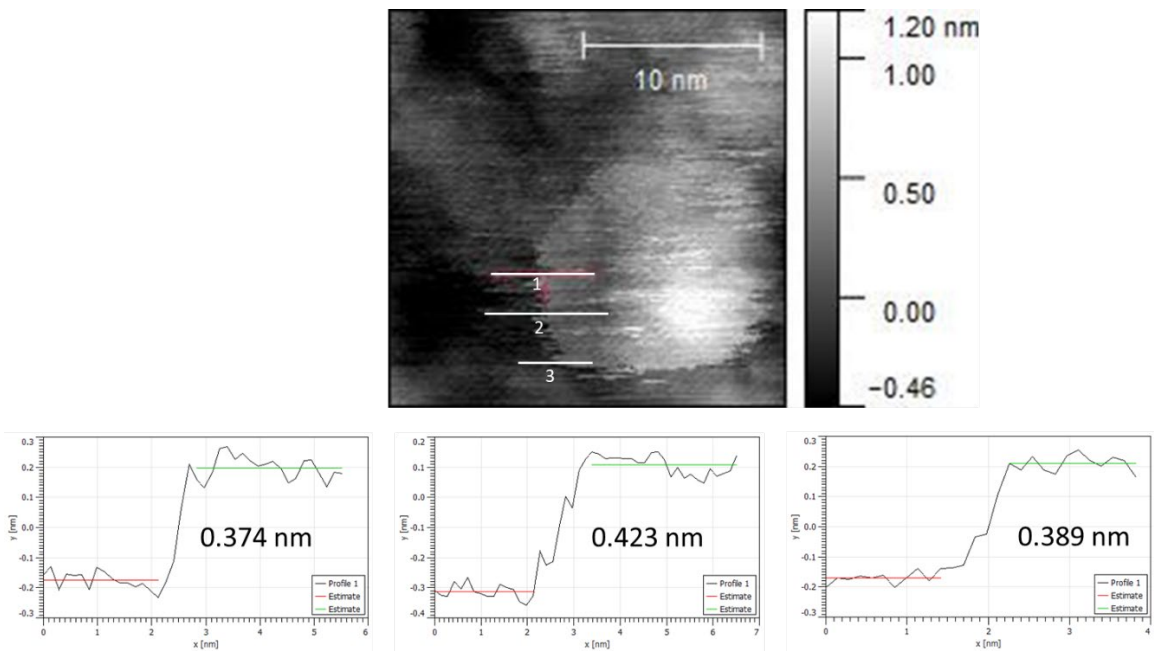

**Figure S14**. High-resolution atomic force microscopy (HR-AFM) image of one individual molecule absorbed on mica surface (top) and its height profiles along the white lines (bottom). Both red and green lines in height profile images are fitted by *Gwyddion*, using zero-order polynomial functions (averaging of the curves).

For this particular single-layer platelet, its mean height and the standard deviation is 3.95±0.20 Å. More platelets were picked out from other images with slightly lower resolutions (due to bigger tip radius, measured with UHF Arrow probes, tip radius around 8 nm) and listed below (**Figure S15**). From these eight platelets, a mean thickness is obtained as 4.04±0.48 Å. Meanwhile, the mean thickness of bilayer clusters is 7.95±0.72 Å (**Figure S16**).

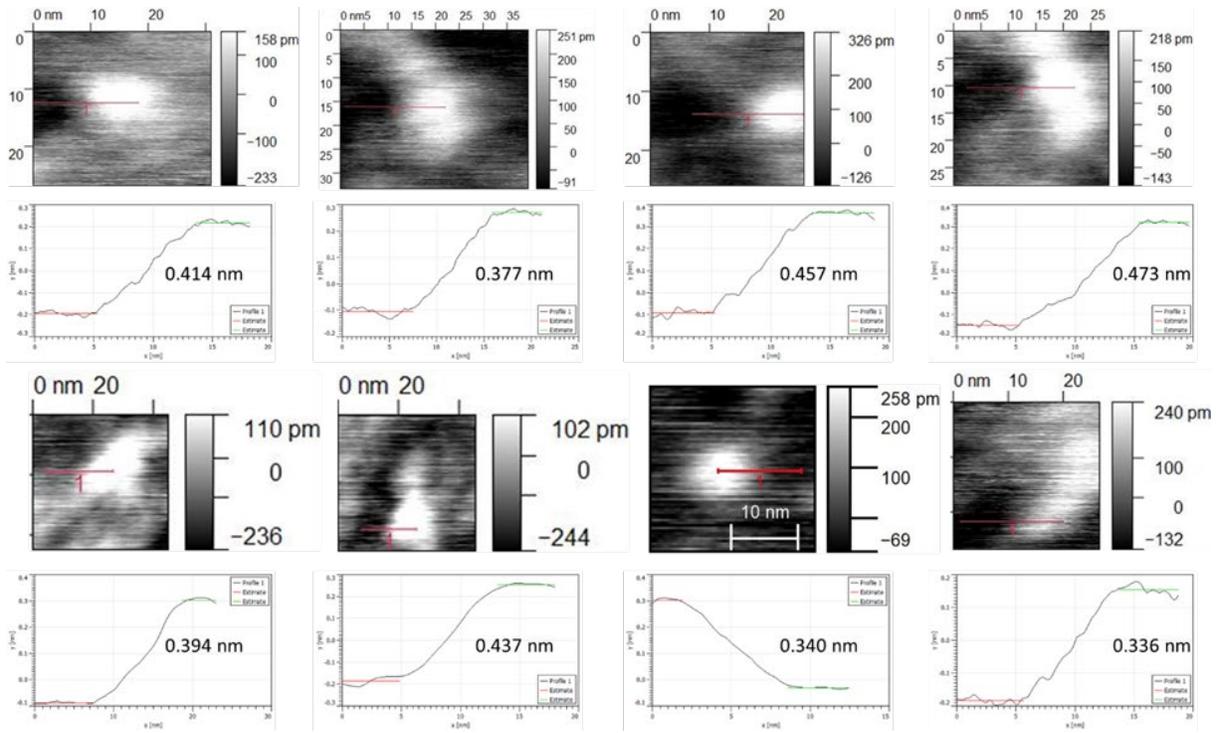

**Figure S15**. Single-layer platelets and their thicknesses. The platelet is chosen nearby a bare mica surface to offer a flat background.

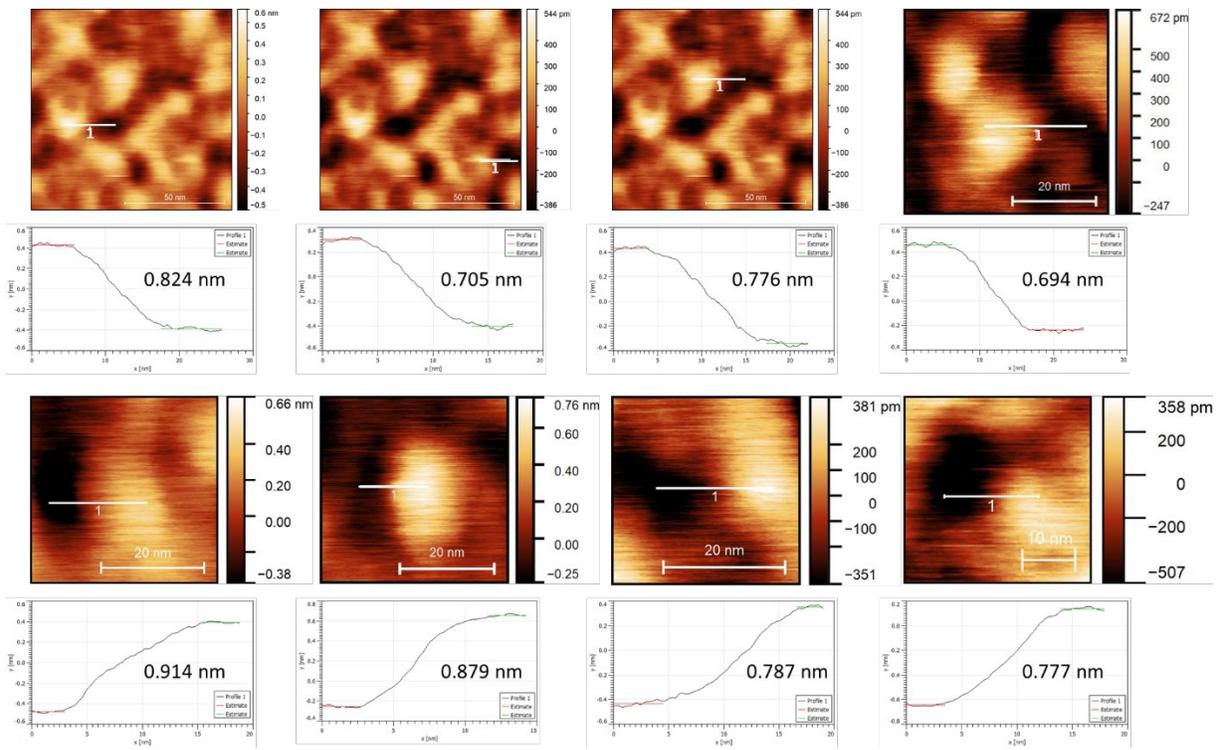

**Figure S16**. Bilayer clusters and their thicknesses.

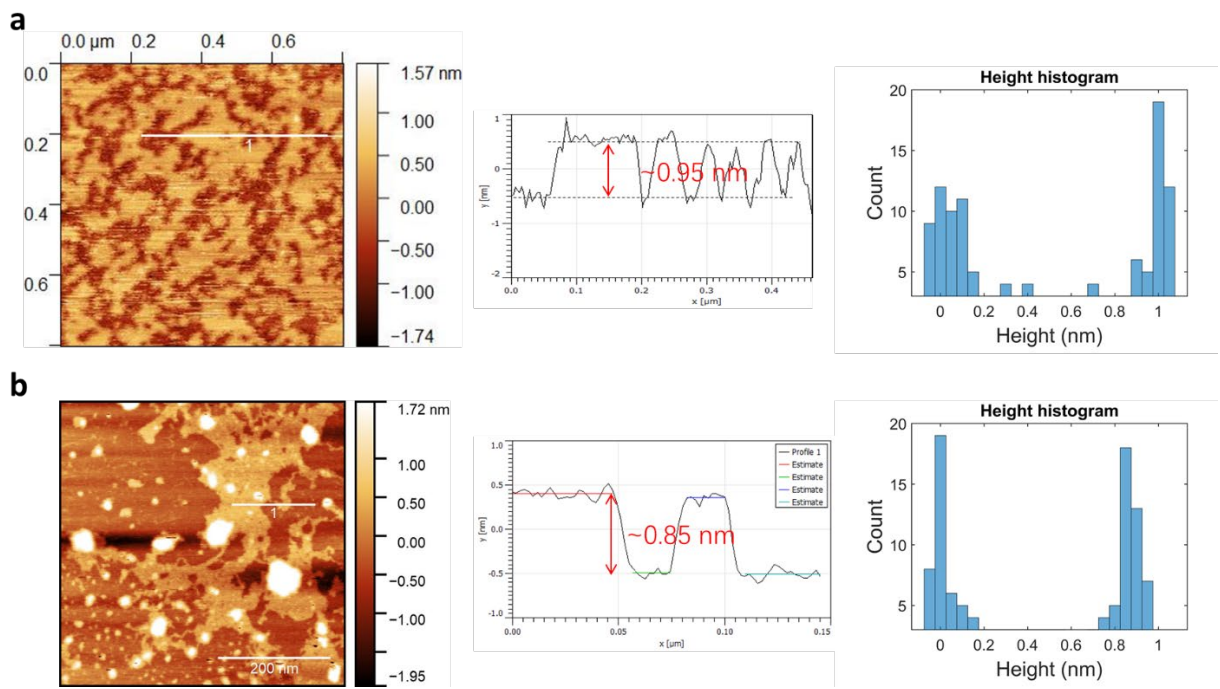

**Figure S17**. High-resolution atomic force microscopy (HR-AFM) image of bilayer nanosheets on mica surfaces (**a** and **b**) and their height profiles along white lines. Height histograms are also shown on the right. **a**, Made from spin-coating. **b**, Made from drop-casting. White dots in **b** are big aggregates formed in the solution phase and then precipitate onto a growing **YZ-2** bilayer film.

When dilute solutions were spin-coated onto a freshly cleaved mica surface, discontinued bilayer films were formed (**Figure S17a**), as an intermediate state to larger films observed in **Figure S18**. Interestingly, if drop-coating is employed instead of spin-coating, the resulting bilayer films are larger and less uniform (**Figure S17b**). We also observed big aggregates on top of those bilayer films. We propose that during the drying process in drop-casting, individual molecules have more time to find existing bilayers on the substrate and thus stitch them together. Meanwhile, as the solution concentration gets larger, molecules also bind with each other in the solution phase, forming big aggregates that precipitate in the end.

In **Figure S18a** and **b**, a layer of nanosheet was formed on mica as a bottom layer, then additional nanosheets start to grow on top of them. In some cases, one layer can grow on two distinct layers with different heights, just as shown in the middle of **Figure S18a**.

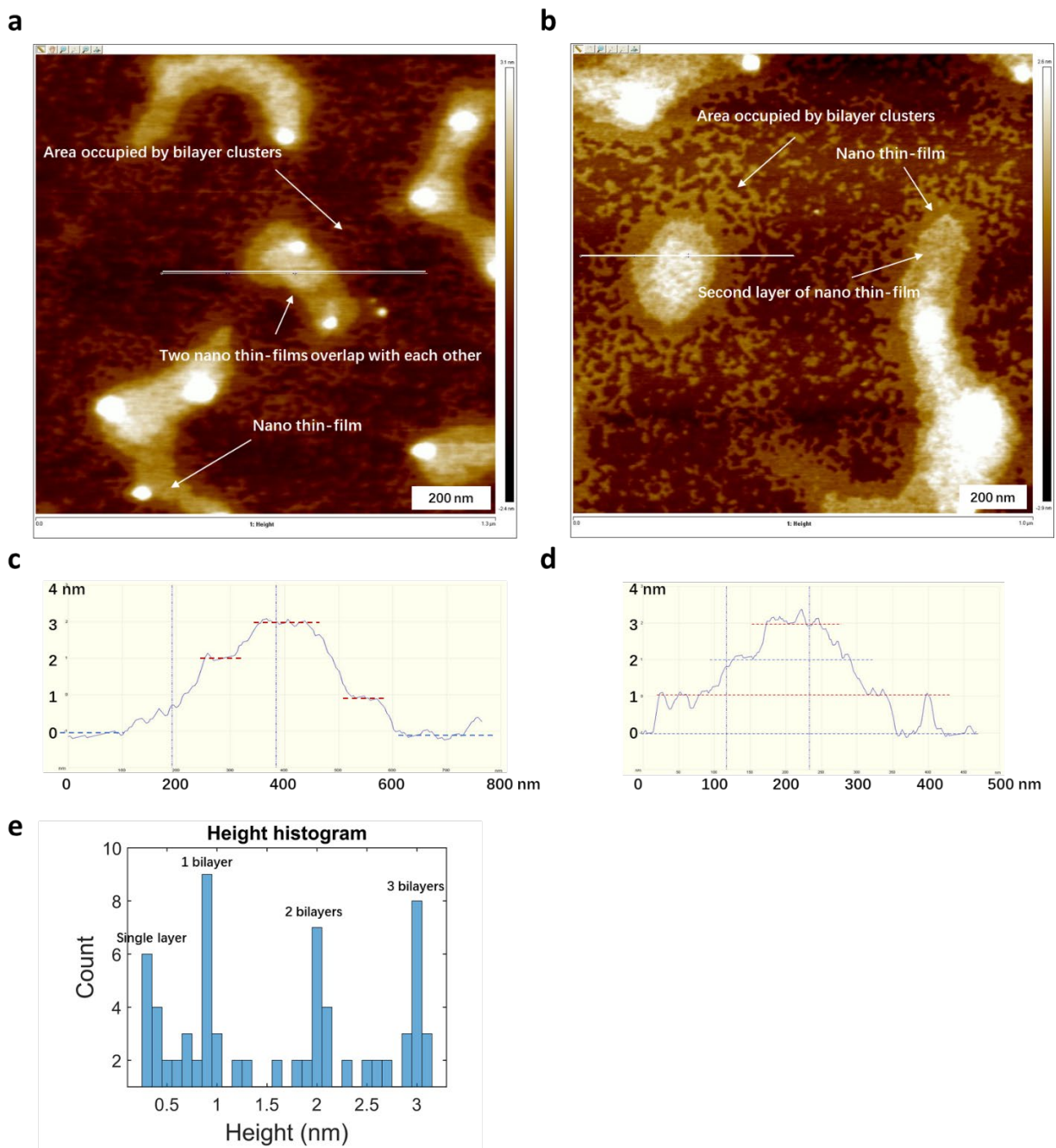

**Figure S18**. High-resolution atomic force microscopy (HR-AFM) images of bilayer nano thin-films on mica surface (**a** and **b**) and their height profiles along white lines (**c** and **d**). **e**, Height histogram of all points in the height profile **c**.

The height histograms in **Figure S17** show a bilayer thickness of around 0.90 nm and its multiples showing at around 2 and 3 nm (**Figure S18**). The discrepancy of thickness in bilayer nanosheets and stacked bilayer nanosheets may stem from a solvent layer intercalated in the stacked nanosheets.

**Transmission Electron Microscopy (TEM) Characterization**

**Preparation of TEM samples**: **YZ-2** powder was mixed with MeOH (0.1 mg in 1 mL) and sonicated for 1 min. The dilute mixture was then drop-casted onto a lacey carbon/Cu TEM grid. TEM characterization was conducted after drying.

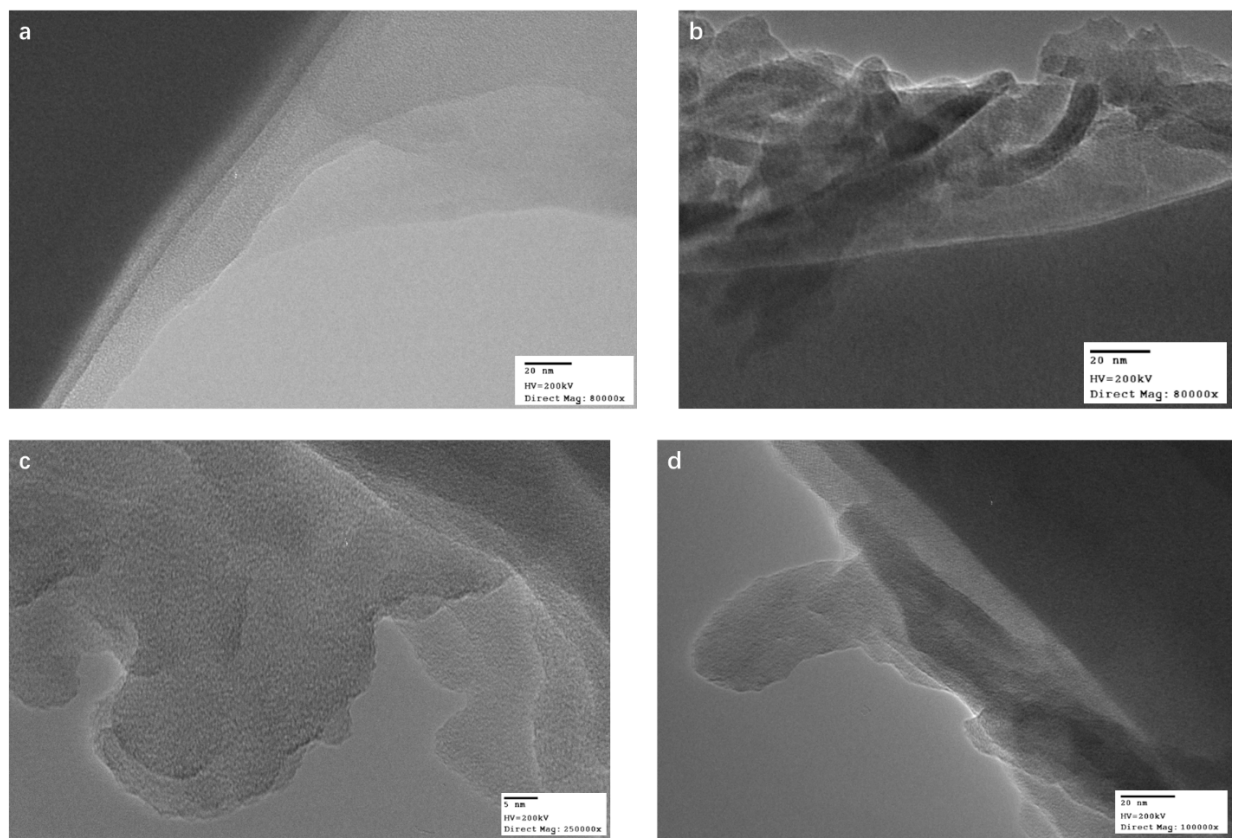

**Figure S19**. Transmission Electron Microscopy (TEM) images of layered structures at the edge of liquid exfoliated **YZ-2** powder

Terrace nanostructures were observed at the edge of liquid exfoliated powders, although the thickness of each step is unknown.

# Film Characterization

**Preparation of YZ-2 thin films: YZ-2** powder was allowed to dissolve in different amounts of trifluoroacetic acid (TFA), forming clear homogenous solutions. Spin-coating of those solutions onto clean $SiO_2$/Si wafers offers flat and uniform thin films.

**Note**: To get uniform thin films on $SiO_2$/Si wafers, substrates have to be pre-cleaned with acetone and isopropanol using a bath sonicator. Dust and contaminations will lead to imperfection and discontinuity.

**Film thickness measurement**: Prepare **YZ-2** solutions with different concentrations (0.5, 1, 2, 5, 10, and 15 mg/mL in TFA). Prepare thin films by spin-coating (2000 rpm, 1 min) onto square substrates (length around 1.5 cm). Each concentration was repeated four times. Make scratches with a fine needle and then measure the film thickness using an AFM at scratches (**Figure S20**). Each sample was measured at five different places to get statistics. Alternatively, the film thickness can be also measured by TEM (**Figure S25**) and SEM (**Figure S26** & **S27**). In most cases, AFM is used because it is more facile.

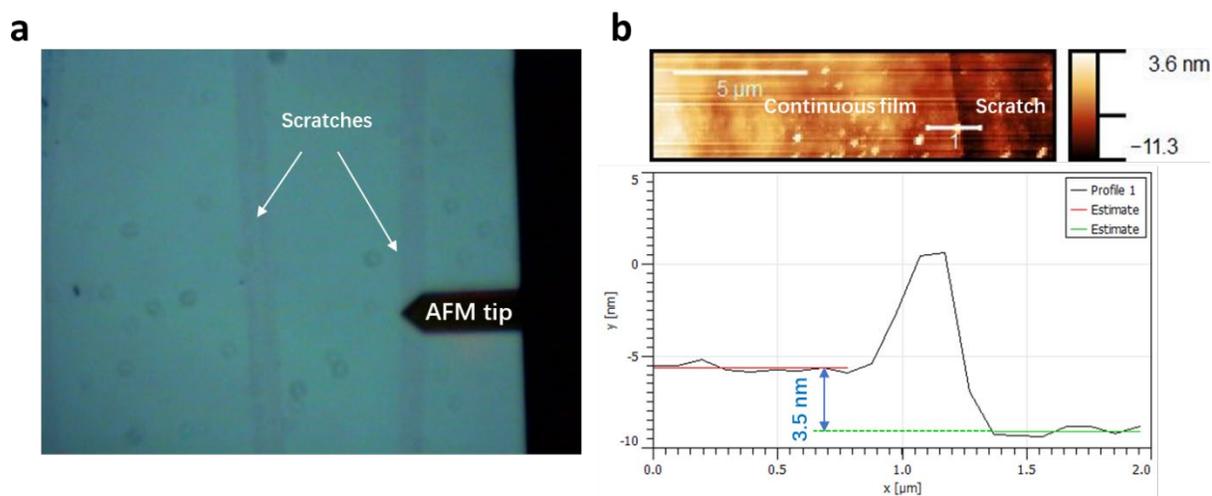

**Figure S20**. **a,** Optical micrograph of thickness measurement. Black dots are dirt on the optical lens. Scratches were made with a fine needle. **b,** Topological image of a thin-film and its thickness. White dots on the thin-film are residual materials made by scratching.

| Concentration: 15 mg/mL | | | | |
|---|---|---|---|---|
| | Sample | | | |
| | A | B | C | D |
| Place 1 | 139 | 119 | 136 | 115 |
| Place 2 | 152 | 107 | 118 | 112 |
| Place 3 | 147 | 91 | 148 | 150 |
| Place 4 | 100 | 144 | 127 | 148 |
| Place 5 | 109 | 167 | 80 | 103 |
| Average | 129.4 | 125.6 | 121.8 | 125.6 |
| Error | 20.94373 | 26.97851 | 23.13785 | 19.52025 |
| | | | | |
| Average (Total) | | 125.6 | | |
| Error (Total) | | 22.97694497 | | |

| Concentration: 10 mg/mL | | | | |
|---|---|---|---|---|
| | Sample | | | |
| | A | B | C | D |
| Place 1 | 80 | 87 | 70 | 80 |
| Place 2 | 82 | 84 | 77 | 79 |
| Place 3 | 78 | 85 | 87 | 78 |
| Place 4 | 86 | 84 | 84 | 75 |
| Place 5 | 72 | 81 | 79 | 66 |
| Average | 79.6 | 84.2 | 79.4 | 75.6 |
| Error | 4.630335 | 1.939072 | 5.885576 | 5.083306 |
| | | | | |
| Average (Total) | | 79.7 | | |
| Error (Total) | | 5.541660401 | | |

| Concentration: 5 mg/mL | | | | |
|---|---|---|---|---|
| | Sample | | | |
| | A | B | C | D |
| Place 1 | 42 | 39 | 38 | 41 |
| Place 2 | 39 | 38 | 40 | 39 |
| Place 3 | 41 | 37 | 38 | 41 |
| Place 4 | 41 | 39 | 41 | 41 |
| Place 5 | 43 | 41 | 37 | 38 |
| Average | 41.2 | 38.8 | 38.8 | 40 |
| Error | 1.32665 | 1.32665 | 1.469694 | 1.264911 |
| | | | | |
| Average (Total) | | 39.7 | | |
| Error (Total) | | 1.676305461 | | |

| Concentration: 2 mg/mL | | | | |
|---|---|---|---|---|
| | Sample | | | |
| | A | B | C | D |
| Place 1 | 17 | 15 | 15 | 17 |
| Place 2 | 17 | 16 | 17 | 18 |
| Place 3 | 14 | 14 | 15 | 13 |
| Place 4 | 16 | 17 | 15 | |
| Place 5 | 14 | 16 | 17 | 14 |
| Average | 15.6 | 15.6 | 15.8 | 15.5 |
| Error | 1.356466 | 1.019804 | 0.979796 | 2.061553 |
| | | | | |
| Average (Total) | | 15.63157895 | | |
| Error (Total) | | 1.384520678 | | |

| Concentration: 1 mg/mL | | | | |
|---|---|---|---|---|
| | Sample | | | |
| | A | B | C | D |
| Place 1 | 8 | 8 | 10 | 10 |
| Place 2 | 8 | 7 | 9 | 9 |
| Place 3 | 8 | 7 | 8 | 8 |
| Place 4 | 9 | 8 | 8 | 8 |
| Place 5 | 7 | 9 | 8 | 8 |
| Average | 8 | 7.8 | 8.6 | 8.6 |
| Error | 0.632456 | 0.748331 | 0.8 | 0.8 |
| | | | | |
| Average (Total) | | 8.25 | | |
| Error (Total) | | 0.829156198 | | |

| Concentration: 0.5 mg/mL | | | | |
|---|---|---|---|---|
| | Sample | | | |
| | A | B | C | D |
| Place 1 | 5 | 4 | 4 | 3 |
| Place 2 | 4 | 5 | 4 | 6 |
| Place 3 | 4 | 4 | 4 | 5 |
| Place 4 | 4 | 5 | 3 | 4 |
| Place 5 | 3 | 4 | 5 | 4 |
| Average | 4 | 4.4 | 4 | 4.4 |
| Error | 0.632456 | 0.489898 | 0.632456 | 1.019804 |
| | | | | |
| Average (Total) | | 4.2 | | |
| Error (Total) | | 0.748331477 | | |

**Figure S21**. The thickness of samples with different spin-coating concentrations

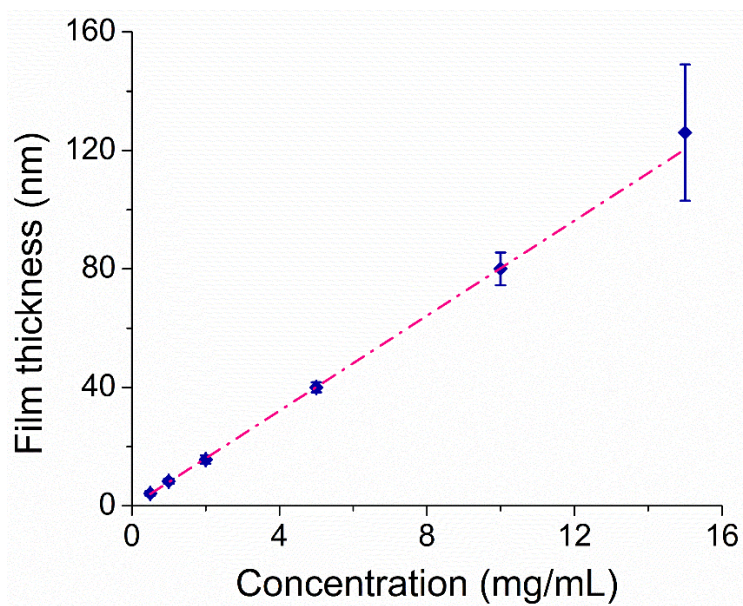

**Figure S22**. Thickness-concentration dependence of spin-coated films on SiO$_2$-covered (300 nm) silicon wafers. Standard deviations of the thickness are shown as error bars.

**Top view of thin films and surface roughness measurement:** The surface topology and roughness of spin-coated thin films were measured by a *Cypher S* AFM from *Oxford instruments* and analyzed with *Gwyddion* or *Cypher*.

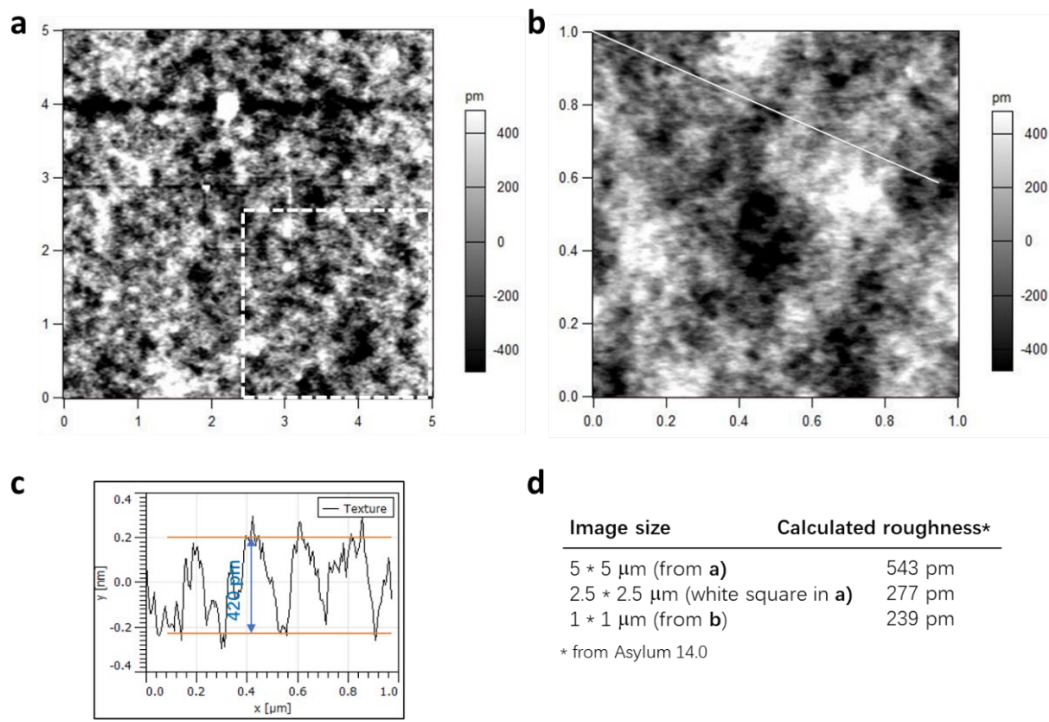

**Figure S23.** AFM images of spin-coated **YZ-2** film surfaces (**a** and **b**). **c,** The texture (height) information along the white line in **b**. **d**, Calculated roughness from selected areas.

All those spin-coated films have super-flat surfaces. Their roughness usually ranges from 300-400 pm over a 5*5 μm area, similar to a commercial ultra-flat silica wafer. In **Figure S23a**, there is a white dot in the upper middle substantially contributing to the surface roughness (543 pm). We attribute it as multiple layer laminates or a 3D amorphous molecule. In a zoomed area (**Figure S23b**), one can easily figure out four layers of molecules from their grayscale and the height profile in **Figure S23c** shows a step of ~420 pm, which corresponds to a single-molecular thickness.

The HR-AFM characterization of thin-films also offers some topologic information. However, unlike measurements of single platelets on the mica substrate, height images of thin-films are vague, probably due to the soft nature of 2D molecules and their imperfect (random) stacking (**Figure S24a**). Fortunately, we found that the image contrast can be dramatically enhanced when the tip oscillates at higher eigenmode values.[45] Therefore, in dual AC mode, the amplitude-channel AFM image at the second eigenmode value gives a much higher sensitivity to the height variation, showing clear edges for individual molecules (**Figure S24b**). Thus, by just measuring the film surface, we can get a top-layer molecular size distribution which should also reflect the real one because, during the spin-coating process, there is no molecular size preselection involved.

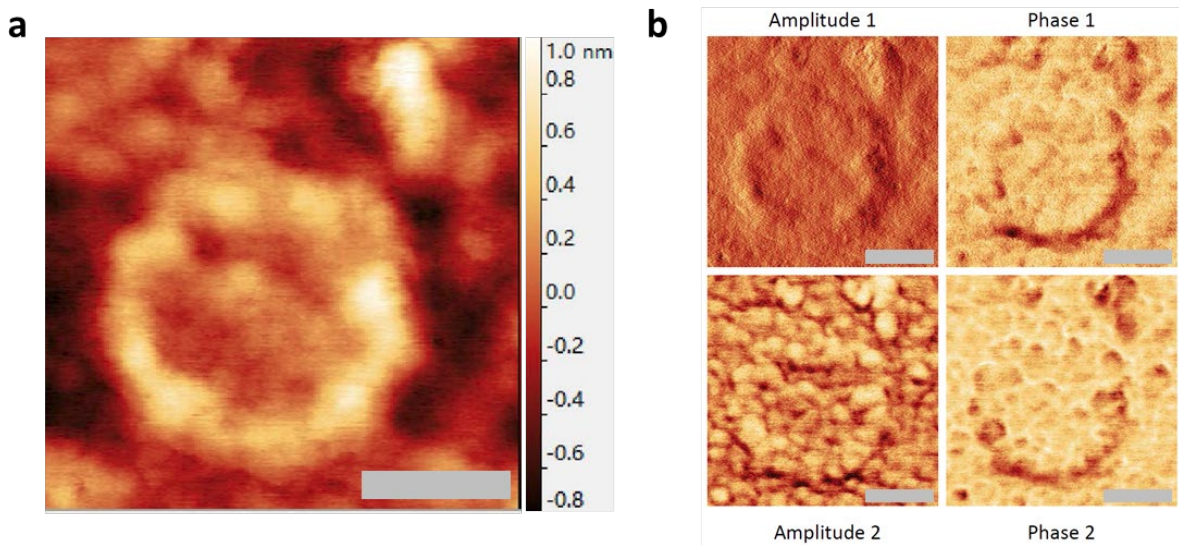

**Figure S24. a,** Original height image of a selected area. Scale bar: 50 nm. **b,** Amplitude and phase images at the first (up) and second (down) eigenmodes. Scale bar: 50 nm.

**The cross-sectional view of thin films and thickness measurement**: The uniformity and thickness of thin films can be also measured by TEM (**Figure S25**) and SEM (**Figure S26**). The TEM sample was prepared by Focused Ion Beam (FIB) cutting of a spin-coated film sample followed by mild argon milling while samples for SEM were directly observed at the fresh-cleaved substrate edge after Au sputtering of a spin-coated film sample. In SEM images, there are some **YZ-2** debris sticking to the fractures but wouldn't affect the thickness measurement.

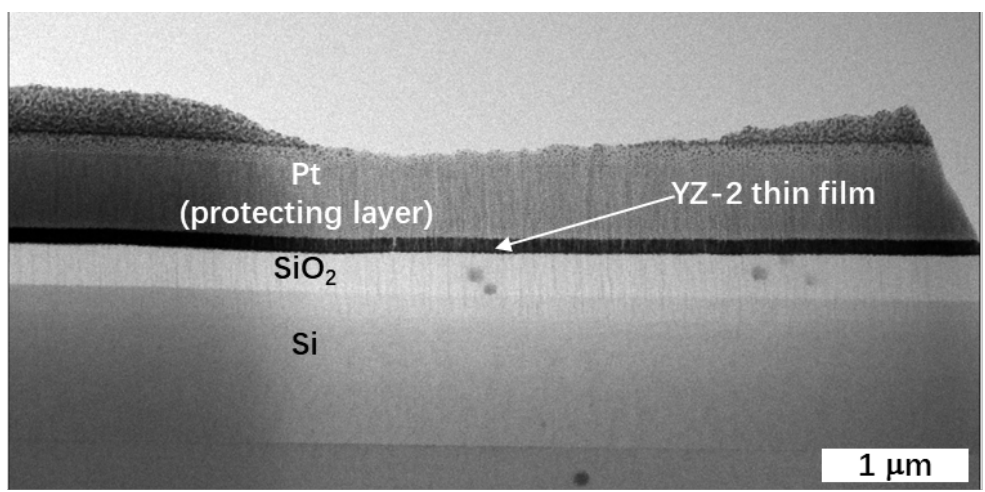

**Figure S25**. A cross-sectional view TEM image of **YZ-2** thin film on a SiO$_2$/Si substrate. Oxide thickness: 300 nm.

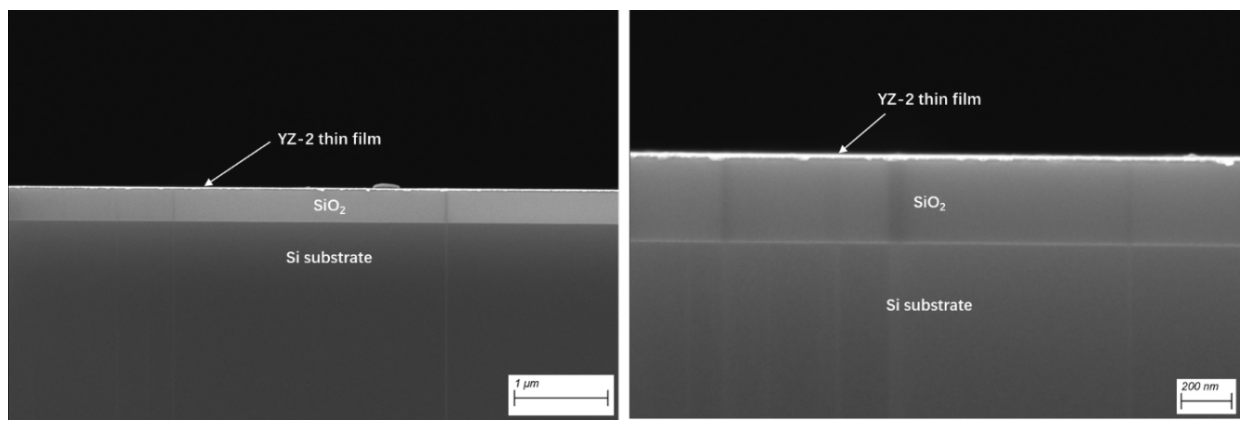

**Figure S26**. Cross-sectional view SEM images of **YZ-2** thin film with different magnifications. Oxide thickness: 300 nm.

**Characterizations of suspended thin films made by drop-casting:** Suspended films were formed when dilute **YZ-2** solution was drop-casted and dried on a holey TEM grid. The desired films were then characterized by SEM (**Figure S27**) and AFM (on a different sample, **Figure S28**).

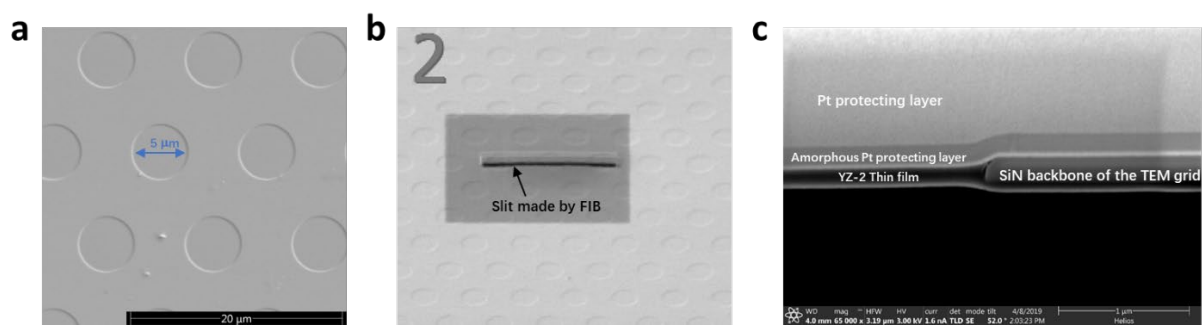

**Figure S27**. Top-view SEM images of suspended films before (**a**) and after (**b**) focused ion beam (FIB) cutting. **c**, Cross-sectional view of FIB cut suspended film. The hole size of the TEM grid: 5 μm.

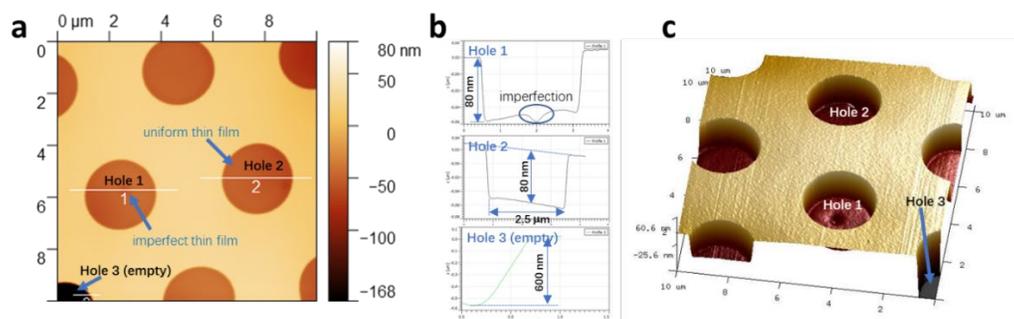

**Figure S28**. AFM topology (**a**), height information (**b**), and 3D reconstruction (**c**) of suspended thin films. The hole size of the TEM grid: 2.5 μm.

In **Figure S28a**, we observed an imperfect thin film (hole #1), containing a dip in the middle of the membrane (**Figure S28b**). There is also an empty hole in the bottom left corner (hole #3). The suspended films first grow along the wall of the grid holes, and then cross it. This observation is similar to that of suspended graphene films and the deep of thin-film implies the strength of substrate-film affinity.[3,34]

## Polarized Photoluminescence (PL) Characterization

**Optical setup**: The whole optical setup is shown in **Figure S29**. A continuous-wave 532 nm laser (Edmund, 35-072) was used for excitation. The incident light traveled through a linear polarizer and a half-wave plate (mounted on a motorized stage) and focused onto the sample using an objective lens (Zeiss, 100x, NA=0.75). The angle of the half-wave plate was adjusted to maximize photoluminescence intensity. Then, the stage was moved a few μm away because **YZ-2** has already photobleached to some extent during the focusing and adjustment of the half-wave plate. The signal was collected with a spectrometer (Princeton Instruments, Acton SpectraPro SP-2150, and PyLon). The excitation power for photoluminescence measurements was 500 μW and the exposure time was 10 seconds. For simple PL measurement, a spectrometer was used for data collection. However, for the polarized PL study, we used an EMCCD camera (Andor, iXon3), which is much more sensitive, to trace a longer time course despite the photobleaching of **YZ-2**. The polarity of the incident light was controlled by rotating the half-wave plate and the PL signal was collected every 5 degrees with 5 seconds exposure time. The excitation power was 2 μW for excitation polarization. All measurements were conducted at room temperature under air.

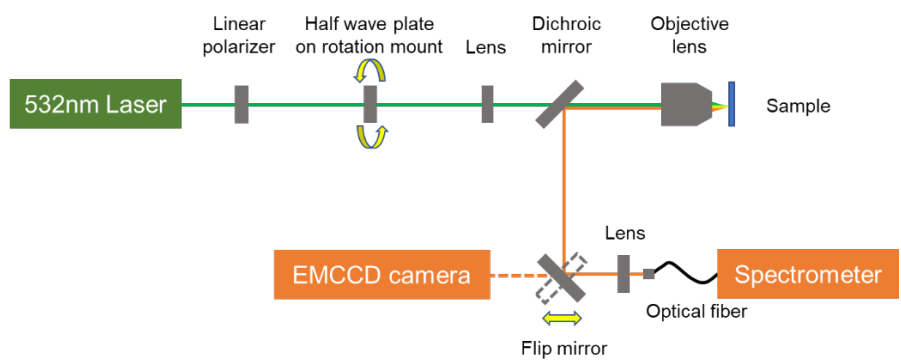

**Figure S29**. Optical setup for Photoluminescence measurements

**Sample preparation and mounting**: First, spin-coat **YZ-2** films onto clean $SiO_2$/Si substrates. For top view measurement, the sample was fixed onto a glass slide with a flat-on orientation. To protect the **YZ-2** film from scratching and contamination, the film side was facing the glass slide (**Figure S30a**). For side view measurement, the sample was cut into two pieces. The freshly cleaved edge was glued onto a glass slide with an edge-on orientation. The film is perpendicular to the glass slide (**Figure S30b**).

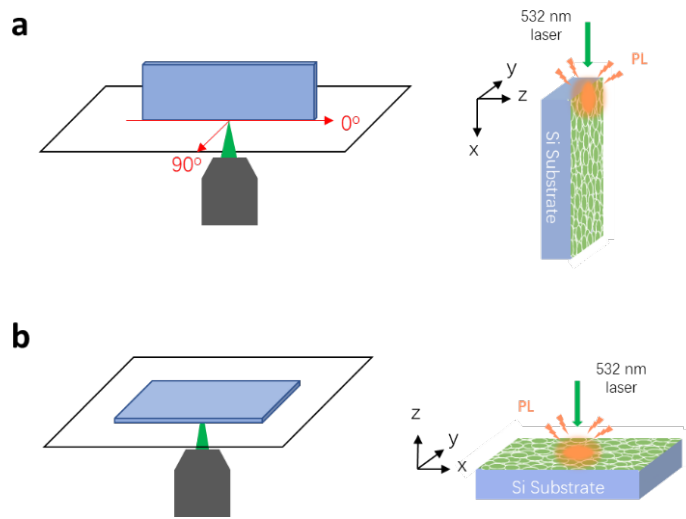

**Figure S30**. Schematic illustration of the experimental setup. **a**, Top view. **b**, Side view.

**Photoluminescence measurement**: Before the polarized PL study, we measured PL spectra of **YZ-2** in bulk powder, solution-phase (10 mg/mL), and different oriented samples (**Figure S31**).

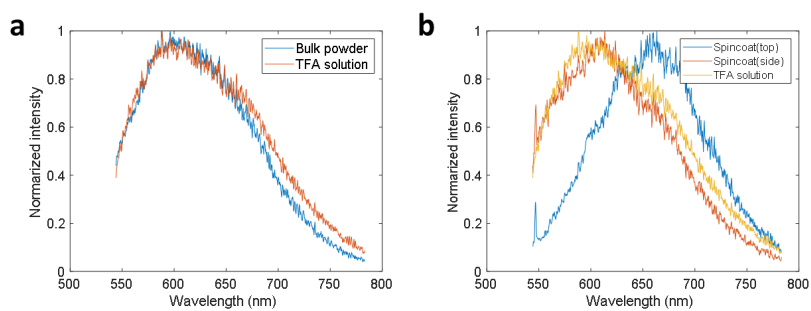

**Figure S31**. **a**, A comparison of Photoluminescence data from bulk powder and TFA solution. **b**, Photoluminescence response from top view sample, side view sample, and TFA solution.

The fact that both bulk powder (condensed state) and TFA solution (highly dispersed state) offer the same PL response indicates that individual molecule excites and emits on its own, without any synergy. The difference between the top view and side view shows the existence of different excitation modes, which are sensitive to the incoming laser pathway.

**Polarized photoluminescence study**: Data was collected using an EMCCD detector. Linear polarization is controlled by rotating a half-wave plate.

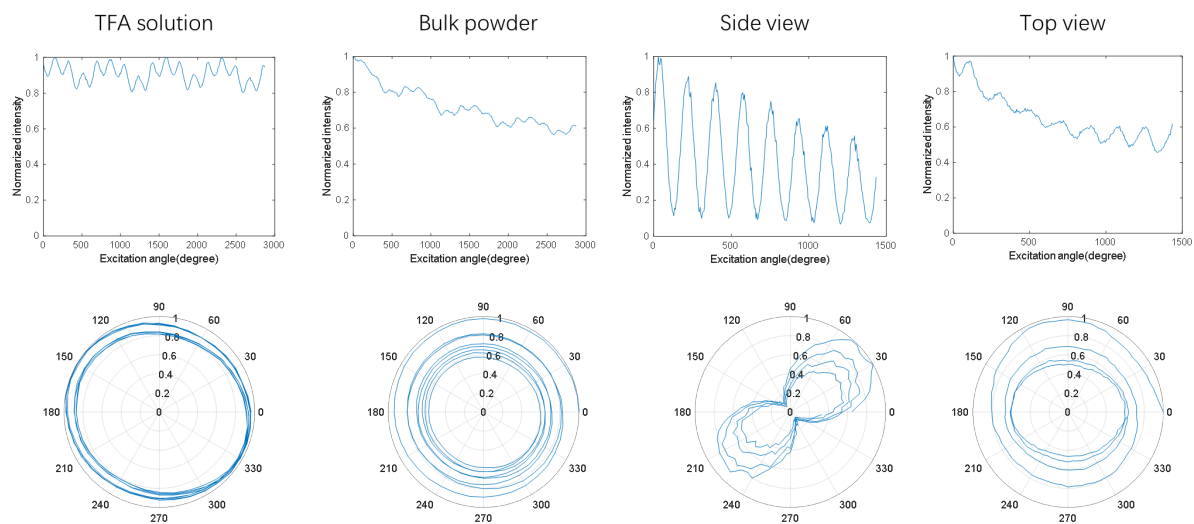

**Figure S32**. Photoluminescence intensity-excitation angle curves (up) and their corresponding polar plots (bottom).

Only the side view sample shows angular dependence, indicating that **YZ-2** molecules are anisotropically aligned in *yz* plane (**Figure S30a**). Likewise, the lack of angular dependence implies that **YZ-2** molecules are isotropically dispersed in *xy* plane (**Figure S30b**). The intensity decrease observed in bulk powder, top view, and the side view is attributed to photobleaching of **YZ-2**, probably originate from laser heating. After fitting and normalizing data with an exponential decay (from first-order kinetics), bleaching corrected data was shown in **Figure S33**.

| Side view | Top view |
|---|---|
| 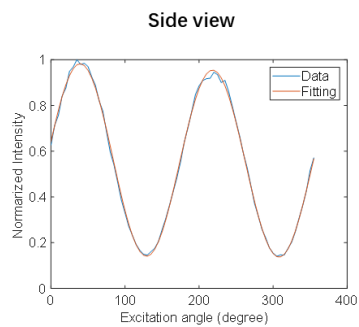 | 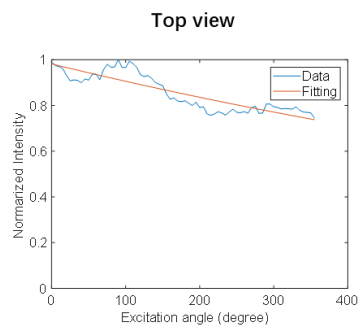 |
| $y = 0.4226 \exp(-1.61*10^{-4} x)*(\sin(0.03512x + 0.1798) + 1.34)$ | $y = 0.9814 \exp(-8.041*10^{-4} x)$ |
| 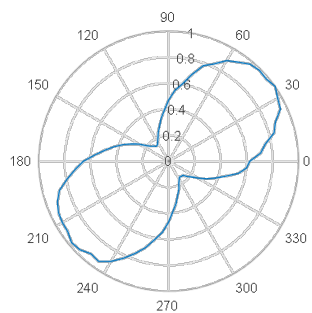 | 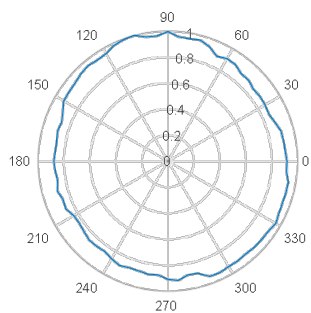 |

**Figure S33**. Fit the PL intensity with first-order decay (up). Normalized polar plots (bottom) of the side view (left column) and top view (right column).

**Grazing-Incidence Wide-Angle X-ray Scattering (GIWAXS) Analysis**

The transferred spin-coated nanofilm (size: around 1.5*1.5 cm; substrate: SiO$_2$/Si) was measured by Dr. Esther Tsai at the beamline 11-BM Complex Materials Scattering (CMS) of National Synchrotron Light Source II (NSLS-II) in the Brookhaven National Laboratory. Thin-film samples were measured at incident angle 0.12 deg with a 200um(H)*50um(V) beam at wavelength λ=0.9184 Å. 2D scattering patterns were obtained using Dectris Pilatus800k at 0.257m downstream the samples. Data analysis was carried out using beamline-developed software SciAnaylsis.[46]

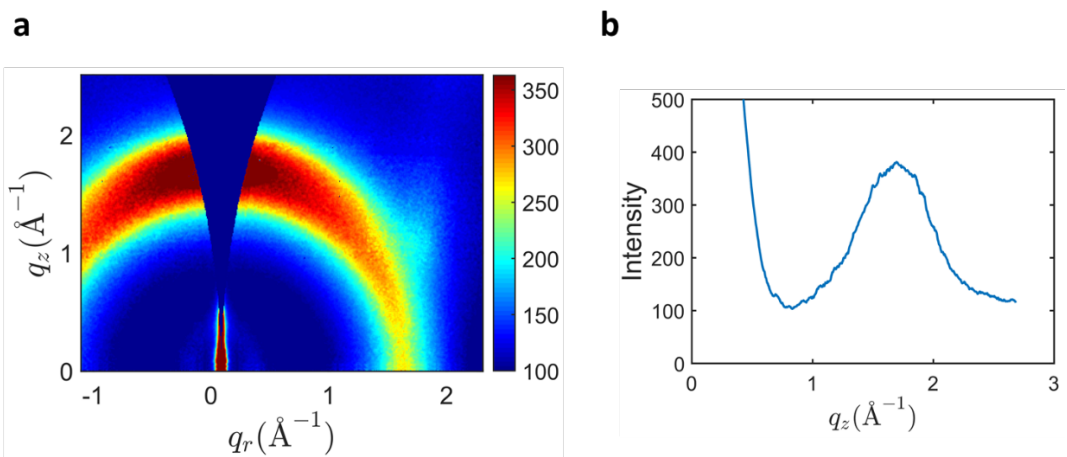

**Figure S34**. **a**, Grazing-incidence wide-angle X-ray scattering (GIWAXS) 2D scattering data of a **YZ-2** thin film obtained from beamline 11-BM of the National Synchrotron Light Source II (NSLS-II). **b**, 1D intensity profile near $q_r$ = 0 A$^{-1}$.

The arc in $q_z$ axis at 1.69 Å$^{-1}$ corresponds to an interlayer spacing of 3.72 Å, close to our observation from PXRD and HR-AFM.

# Chemical Force Spectroscopy

**Experimental design**: For conventional 2D materials such as graphene and *h*-BN, all covalent bonds are constrained in the 2D plane and there is no out-of-plane dipole movement. In **Table S1**, we rotated the two layers of the structures to generate initial structures of possible in-plane bi-layered structures. Taking a **YZ-2** unit cell as the first layer, we placed the second layer in different orientations. We rotated the two layers with respect to each other from 0° to 180° and 3.45 Å away from the first layer. We carried out variable cell relaxation and successfully obtained 22 structures with converged results (see Methods below) and a realistic cell was obtained. The most stable in-plane **YZ-2** structure is shown in **Figure S35a**.

**Table S1**. The optimal lattice parameters (a, b, c, in Å and α, β, and γ in °) interlayer distance (d in Å), and relative total energy of calculated in-plane **YZ-2** structures (in kcal/mol). The relative energy is reported with respect to the structure with the lowest energy of the 22 calculated structures.

| Structures | a (Å) | b (Å) | c (Å) | α (°) | β (°) | γ (°) | d (Å) | Relative energy (kcal/mol) |
|---|---|---|---|---|---|---|---|---|
| 1 | 11.06 | 11.04 | 6.94 | 87.91 | 91.12 | 119.82 | 3.47 | 0.00 |
| 2 | 11.07 | 11.06 | 6.96 | 90.58 | 88.48 | 120.01 | 3.48 | 1.54 |
| 3 | 11.07 | 11.09 | 6.95 | 89.19 | 90.57 | 120.01 | 3.48 | 1.61 |
| 4 | 11.09 | 11.06 | 6.93 | 90.60 | 89.31 | 119.80 | 3.47 | 1.84 |
| 5 | 11.02 | 11.04 | 6.96 | 89.41 | 92.12 | 119.91 | 3.48 | 3.28 |
| 6 | 11.02 | 11.03 | 6.95 | 89.24 | 92.05 | 119.89 | 3.48 | 3.36 |
| 7 | 11.10 | 11.03 | 6.99 | 86.91 | 88.73 | 120.61 | 3.50 | 3.40 |
| 8 | 11.10 | 11.03 | 6.99 | 86.91 | 88.73 | 120.61 | 3.50 | 3.40 |
| 9 | 11.10 | 11.03 | 6.99 | 86.91 | 88.73 | 120.61 | 3.50 | 3.40 |
| 10 | 11.11 | 11.09 | 7.00 | 89.98 | 90.00 | 119.92 | 3.50 | 3.57 |
| 11 | 11.11 | 11.11 | 7.00 | 90.02 | 90.09 | 120.15 | 3.50 | 3.58 |
| 12 | 11.11 | 11.09 | 6.99 | 89.98 | 90.01 | 119.90 | 3.50 | 3.58 |
| 13 | 11.11 | 11.08 | 7.00 | 90.00 | 89.99 | 119.89 | 3.50 | 3.58 |
| 14 | 11.09 | 11.07 | 6.92 | 90.08 | 89.69 | 119.94 | 3.46 | 3.63 |
| 15 | 11.09 | 11.07 | 6.92 | 90.08 | 89.69 | 119.94 | 3.46 | 3.77 |
| 16 | 11.09 | 11.07 | 6.92 | 90.08 | 89.69 | 119.94 | 3.46 | 5.53 |

| | | | | | | | |
|---|---|---|---|---|---|---|---|
| 17 | 11.08 | 11.07 | 7.22 | 90.09 | 90.02 | 119.83 | 3.61 | 6.24 |
| 18 | 11.07 | 11.07 | 7.22 | 90.10 | 89.98 | 119.81 | 3.61 | 8.23 |
| 19 | 11.08 | 11.11 | 7.20 | 89.97 | 90.01 | 120.01 | 3.60 | 8.28 |
| 20 | 11.10 | 11.09 | 7.18 | 89.98 | 90.05 | 119.94 | 3.59 | 8.38 |
| 21 | 11.10 | 11.09 | 7.18 | 90.01 | 89.98 | 119.89 | 3.59 | 8.38 |
| 22 | 11.10 | 11.09 | 7.18 | 89.98 | 90.05 | 119.94 | 3.59 | 8.41 |

In such a flat structure, the interlayer spacing is 3.47 Å, similar to that of graphene[24] and hBN[25] but smaller than our real interlayer spacing (3.72 Å from GIWAXS). This suggests that in the **YZ-2**, amides are not totally flat and pointing outside of the surface, forming interlayer hydrogen bonding. In such a case, one would expect local surface charges and out-of-plane dipoles. This phenomenon has been observed in 1D systems. For instance, in Kevlar, the amide planes and aromatic cores are not within a same plane.[23,47] Instead, all amide bonds tilt to certain degrees, resulting from steric hindrance between amides and ortho substitutions (H) on the benzene rings.

**Density estimation:** For the most stable in-plane structure (structure 1 in **Table S1**), its mass density can be calculated as

$$\rho_m = \frac{\sum_i^{C,N,H,O} n_i M_i}{N_A V_{cell}}$$

where $\rho_m$ is the mass density of the structure, $n_i$ and $M_i$ is the number of elements, i, and molar mass of element i with i representing C, N, H, and O. $V_{cell}$ is the cell volume, and $N_A$ is the Avogadro's number.

A mass density of 1.28 g/cm³ was calculated and shown in **Table S2**. Since the real structure of **YZ-2** has a larger interlayer spacing, the real mass density of **YZ-2** should be smaller than this calculated density. Thus, this density from an entirely in-plane and densely packed structure can serve as a boundary (upper limit) for the real mass density of **YZ-2**.

**Table S2**: The mass density of the predicted in-plane structure.

| Cell Volume($V_{cell}$) | $n_C$ | $n_H$ | $n_N$ | $n_O$ | $\rho_m$ |
|---|---|---|---|---|---|
| 734.29 Å³ | 24 | 12 | 12 | 6 | 1.28 g/cm³ |

Estimated density ratio of steel (structural steel, ASTM A36) and **YZ-2**: 7.85/1.28 = 6.13

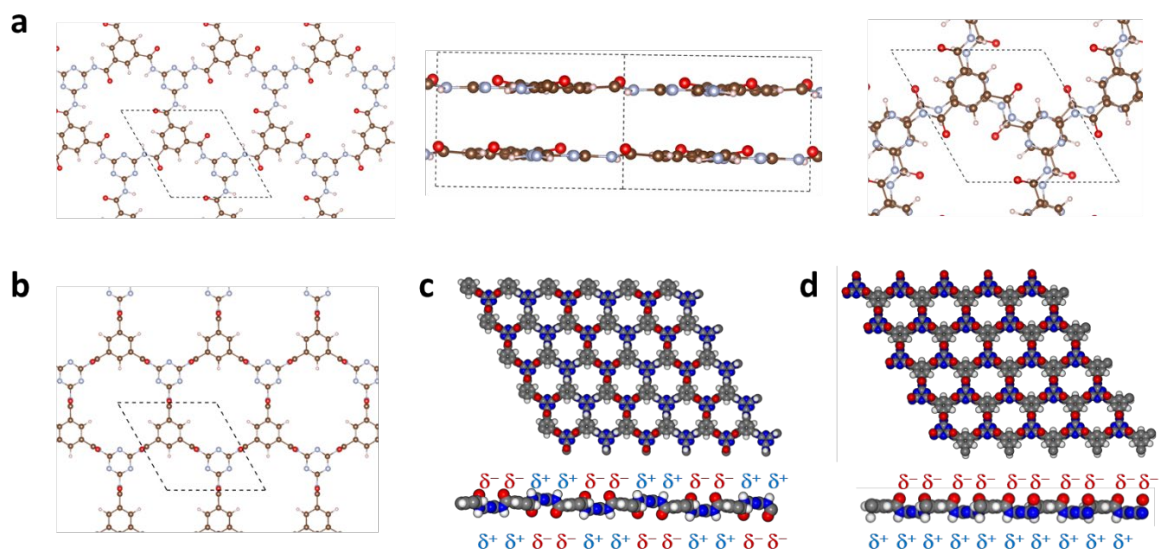

**Figure S35**. **a**, Schematic illustration of an in-plane structure in which all the dihedral angles of amides are 0°. Left: top view of a single layer; middle: side view of a bilayer; right: top view of a bilayer. In this structure, the net surface charge is zero and there is no out-of-plane dipole exists. **b**, Schematic illustration of an out-of-plane structure. The amide orientation is set to 90° for better illustration. **c**, Ball-and-stick model of a trans out-of-plane structure, in which all the amide dipoles are canceled out. **d**, Ball-and-stick model of a cis out-of-plane structure, in which all the amides are pointing to a direction and form a signification dipole across the 2D plane.

We envision that the steric effect mentioned above may also exist in the 2D polyaramid system, despite the strong conjugation interactions (**Figure S35b**). In this scenario, depending on the molecular tacticity, individual charges and dipoles may or may not cancel out. In the former case, the molecule has a symmetric nature and identical surfaces (**Figure S35c**) while in the latter one, the molecule shows a net dipole across the molecular plane, offering two distinct charged surfaces (**Figure S35d**). Although the amide orientations in the real structure are unknown, we exaggerated the dihedral angle of amides and benzene rings to 90 degrees for a better illustration.

To distinguish all above possibilities, we designed a chemical force spectroscopy measurement in which both the sensor (AFM tip) and substrates are chemically modified to selectively bond with different molecular surfaces. Firstly, the $SiO_2$ substrate is known to have an oxygen-rich surface and can hydrogen bond with the NH terminal of the amides (-CO-NH-), leaving the CO-rich molecular surface facing up (**Figure S37a**). By modifying the $SiO_2$ substrate with (3-aminopropyl)triethoxysilane (APTES), the substrate surface is covered with a layer of $NH_2$ residues (**Figure S36b**), which would selectively hydrogen bond with the CO terminal of amides, leaving the NH-rich molecular surface outside (**Figure S37b**). That molecular recognition can be further extended to the next **YZ-2** layer repeatedly, passing from the substrate surface to the film surface. Meanwhile, we coated the AFM tip with a layer of negative charged organic molecules which can "feel" the difference of molecular surfaces (**Figure S36a**).

**Methods**: We used periodic density functional theory (DFT) with the plane-wave code Quantum-ESPRESSO to simulate the theoretical in-plane structure of the **YZ-2**. The plane-wave code, QUANTUM ESPRESSO, was used to determine the lattice constants. The generalized gradient approximation (GGA) with the PBE exchange-correlation functional was applied in conjunction with D3 semi-empirical dispersion to determine lattice constants. Norm-conserving pseudopotentials were employed for all elements with 2s and 2p states in the valence (i.e., for C, N, O) or 1s states in the valence (i.e., H). A plane wave kinetic energy cutoff of 680 eV was employed along with default energy convergence and force thresholds of $10^{-4}$ eV and $10^{-3}$ eV/A, respectively. Lattice constants were determined by relaxing ionic positions and using variable cell relaxation with the same Monkhorst-Pack k-point sampling of 8 × 8 × 4.

Chemical force mappings were performed on a *Bruker Veeco Multimode 8* instrument. To eliminate the influence of the surface water layer and contaminations, all measurements were done under fluid mode using a liquid cell. Deionized water was used as an experimental medium. Moreover, to further minimize the influence of different probes, all data, including substrate controls, were obtained in one measurement without changing probes.

**Modification of AFM probes**: To a 20 mL vial, sodium 3-mercapto-1-propanesulfonate (20 mg) was dissolved in 5 mL EtOH to offer a dilute solution. Before use, each AFM probe (gold coated, NPG-10 from *Bruker AFM Probes*) was immersed in this solution for 10 h (**Figure S36a**).

**Sample Preparation:** APTES-SiO$_2$ substrates were synthesized by immersing clean SiO$_2$ substrates in a (3-aminopropyl)triethoxysilane (APTES) solution (100 mg in 4 mL EtOH) for 10 h (**Figure S36b**). Samples were made by spin-coating (2000 rpm, 1min) **YZ-2** solution (2 mg/mL) onto SiO$_2$ and APTES-SiO$_2$ substrates (**Figure S37**).

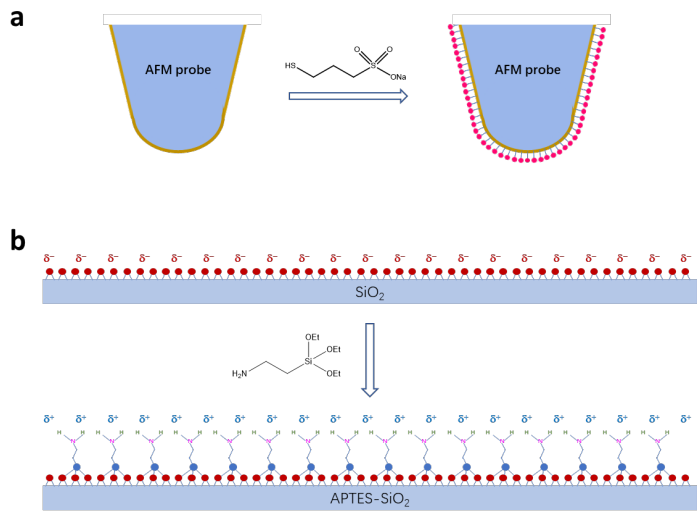

**Figure S36. a**, Modification of gold-coated AFM probes. **b**, Surface coating of SiO$_2$ substrates.

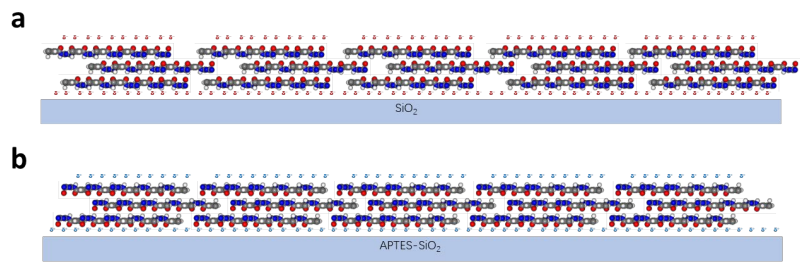

**Figure S37**. Surface recognition leads to different film surfaces. **a**, **YZ-2** film on SiO$_2$ substrate. **b**, **YZ-2** film on APTES-SiO$_2$ substrate.

Surface adhesions of **YZ-2** films on both SiO$_2$ and APTES-SiO$_2$ substrates were measured and shown in **Figure S38**. The significant difference in adhesion force clearly indicates **YZ-2** molecule has a highly asymmetric nature in which amides are not evenly distributed across the molecular plane.

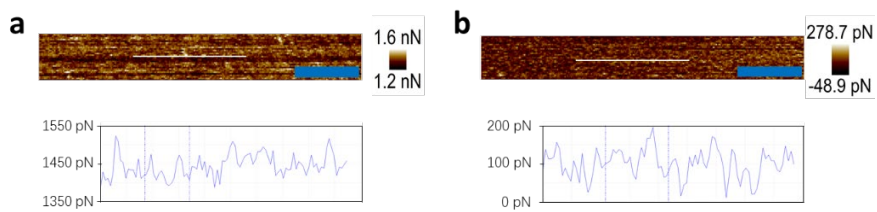

**Figure S38**. Chemical force mapping of **YZ-2** films on SiO$_2$ (**a**) and APTES-SiO$_2$ (**b**) substrates. Adhesion profiles along the white lines are also given in the bottom. Scale bar, 100 nm.

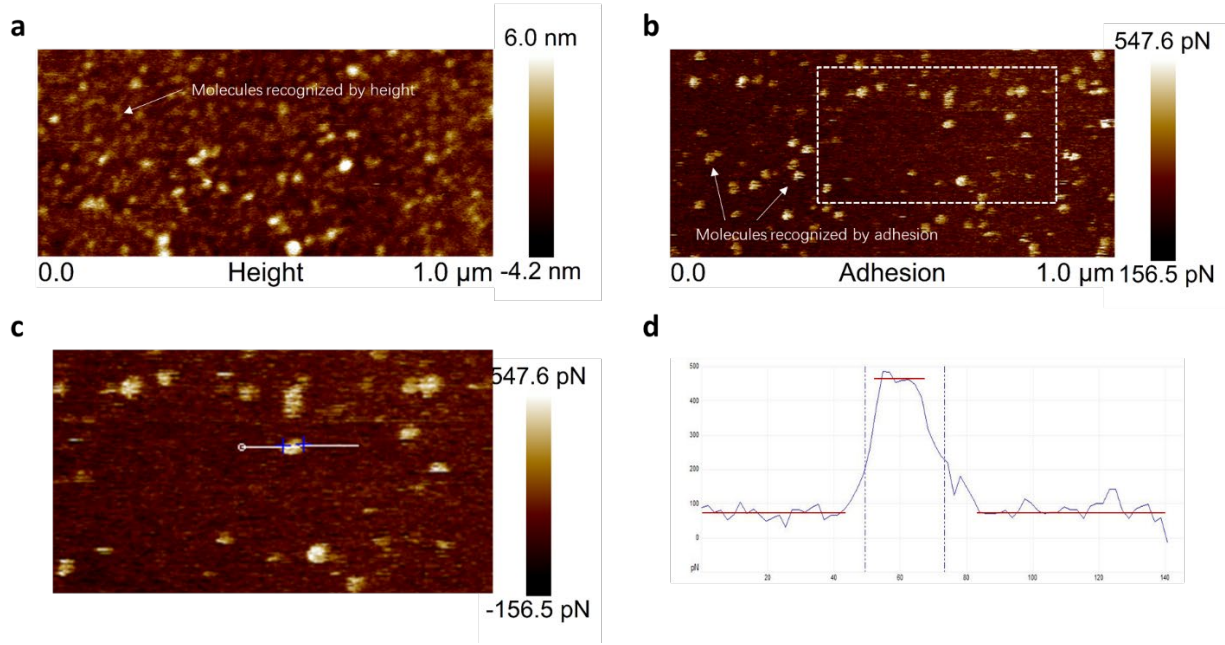

**Figure S39**. Correlated images from height channel (**a**) and adhesion channel (**b**). **c**, Magnified adhesion image from the white rectangle in **b**. **d**, Adhesion profile along the white line in **c**.

Height image and adhesion image in **Figure S39** (**a** and **b**) cannot overlay with each other, which implies that the changing of adhesion is not caused by height variation. We therefore attribute the circular dots in the adhesion image to flipped molecules. This flipping may originate from imperfect APTES coating (**Figure S40**).

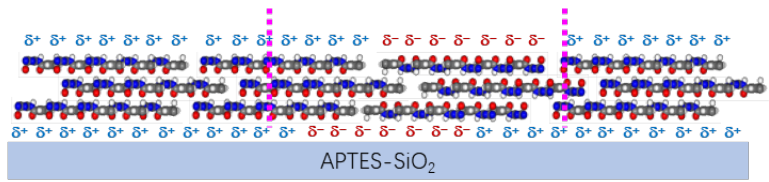

**Figure S40**. An imperfect coating may lead to local molecular flipping

Another evidence for local molecular flipping is similar size distributions from both the height channel and adhesion channel (**Figure S41**).

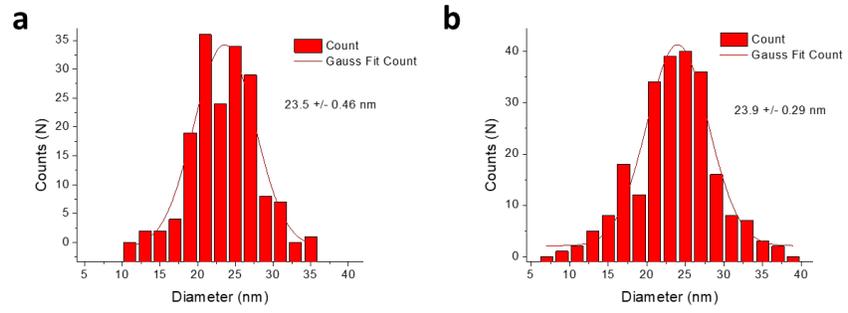

**Figure S41**. Lateral size distributions from height image (**a**) and adhesion force map (**b**)

# Mechanical Measurements

The 2D Young's modulus ($E^{2D}$) and yield strength ($\sigma^{2D}$) were measured at MIT and verified by Army Research Laboratory. The scrolled fiber test was done at MIT.

**AFM Nanoindentation**

**Materials:** A Cypher AFM is used for nanoindentation. Si holey substrates were made by photolithography. Diamond-like spheric probes (Biosphere NT_B50_v0010 and NT_B100_v0010) were purchased from Nanotool AFM Probes. Tip radius information (50±5 nm and 100±10 nm) were obtained from the vendor. Their spring constants and deflection inVOLs were manually calibrated on the same AFM and then verified with a flat, thick polystyrene film on Si/SiO$_2$ wafer. Films can be either formed in situ or transferred onto the substrates. Their thicknesses were obtained by taking an average at 7 different positions nearby the place of interest.

**Film transferring method**: Due to the strong interaction between the polar **YZ-2** molecules and SiO$_2$ surface, one may not peel the **YZ-2** nanofilm off from a SiO$_2$ substrate without destroying it. However, we can pre-lay a supporting polymer layer beneath and peel the whole composite nanofilm off from the substrate. The composite film can be trimmed, cut, and transferred to whatever substrates (different materials, different shapes) after flipping, with the polymer layer facing outside. Subsequent organic solvent washing removes the supporting polymer, leaving the **YZ-2** film alone on the new substrate. This method can easily handle nanofilms up to 6 inches and can afford free-standing membranes across large wells up to 70 μms (**Figure S42**).

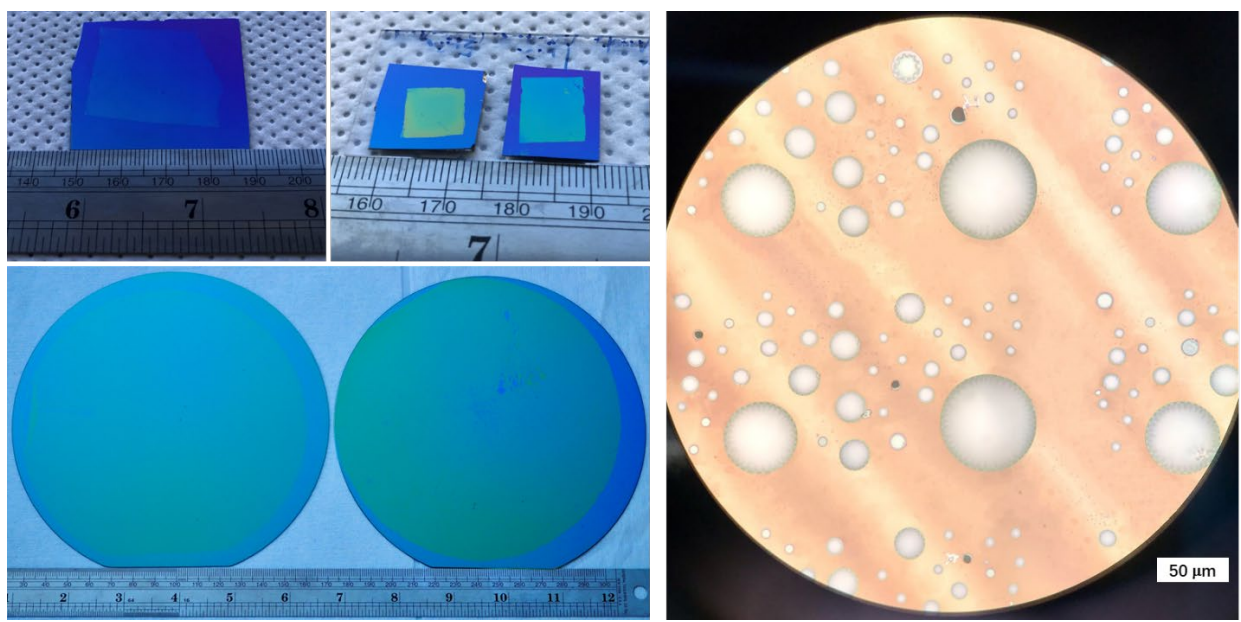

**Figure S42**. Transferred films with different sizes. Left: on SiO$_2$/Si wafers; right: on a holey substrate.

The homogeneity of transferred films was demonstrated by AFM, measured at the film edges, near which cracks, wrinkles, and folds are observed (**Figure S43a&c**). Again, the film shows a very smooth surface and a steady thickness of 7 nm (**Figure S43b**).

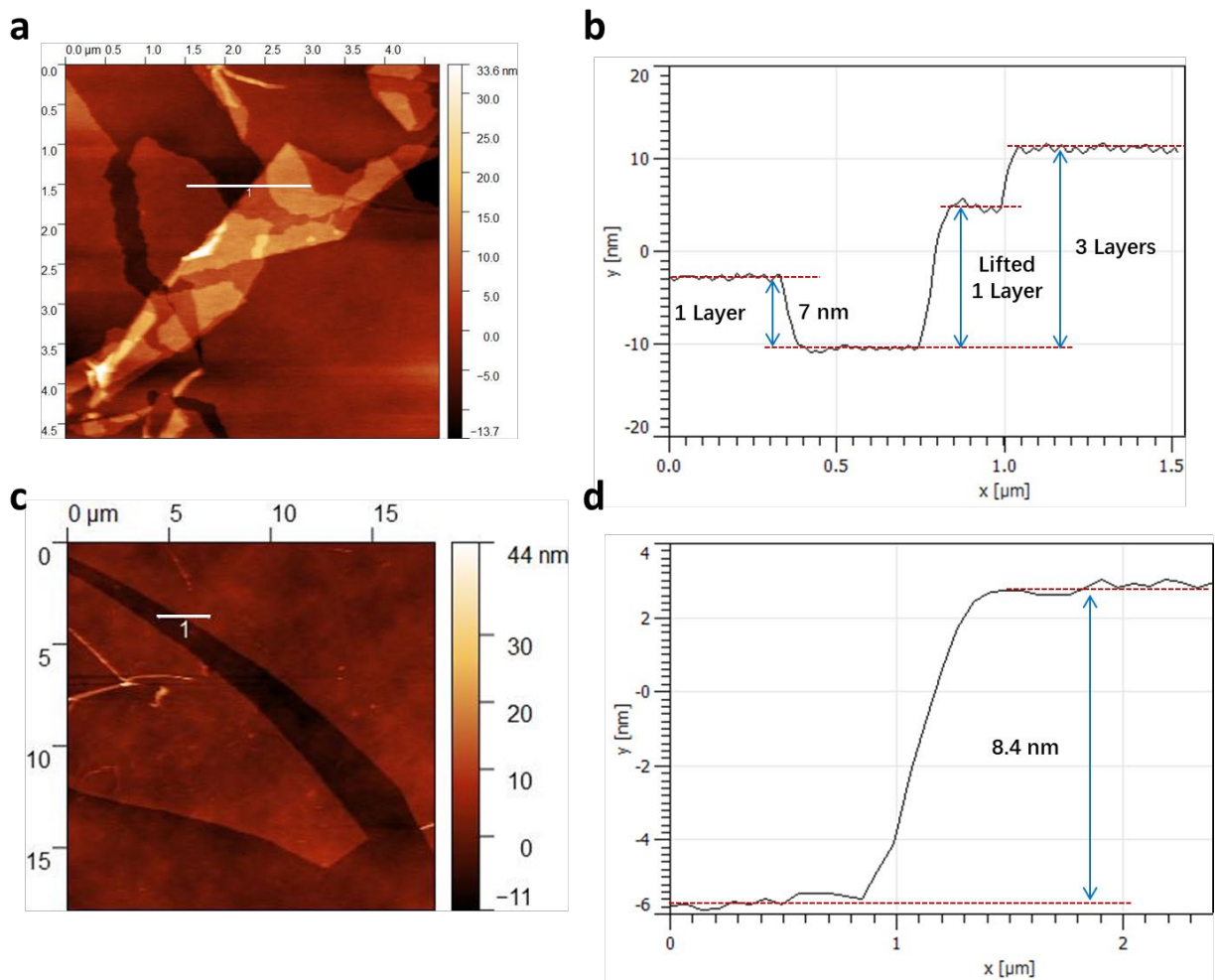

**Figure S43**. AFM images of two transferred **YZ-2** films on $SiO_2$/Si substrates (**a**, **c**) and their height profiles (**b**, **d**) along white lines in **a** and **c**. Measured at cracks near the film edges.

**Nanoindentation method**: The AFM nanoindentation is performed on suspended membranes sitting on holey substrates (**Figure S44**). Two samples with different film thicknesses (33.9 nm and 12.8 nm, measured from 7 different places nearby) were prepared and tested. Nanoindentation was performed on 14-um and 24-um membranes, using two types of probes (synthetic diamond, spheric tips with a radius of 50±5 nm or 100±10 nm). Probes were calibrated manually with a bare substrate surface and verified with a thick, flat polystyrene film.

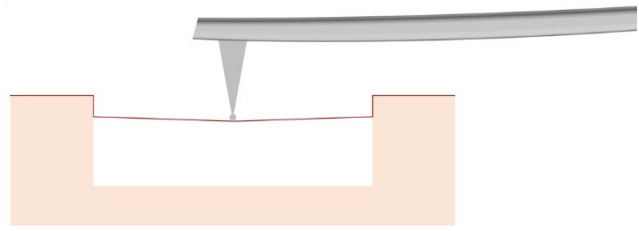

**Figure S44.** Schematic illustration of nanoindentation with a spheric probe on a suspended nanofilm.

120 membranes were measured in total. For each membrane, multiple force-displacement curves with different trigger forces were obtained at the membrane center after releasing gas captured underneath the film. All force curves were plotted in one scheme and their elastic region (the area that curves overlap) was picked out and fitted with the following equation to get the 2D Young's modulus ($E^{2D}$) (**Figure S45**).[3]

$$F = \sigma_0^{2D}(\pi a)\left(\frac{\delta}{a}\right) + E^{2D}(q^3 a)\left(\frac{\delta}{a}\right)^3$$

Where $\sigma_0^{2D}$ is the film pretension, $\delta$ is the deflection at the center point, $a$ is the membrane diameter, and $q$ is a dimensionless constant calculated from Poisson's ratio $\nu$.

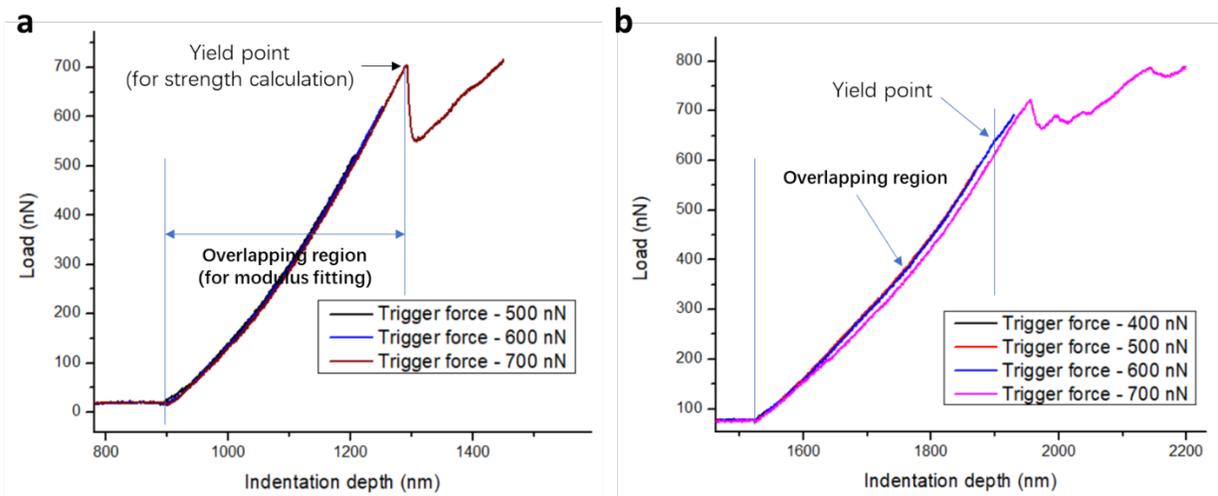

**Figure S45.** Determining the elastic region and yield point for force-displacement curves. **a,** Force curves with a clear yield point. **b,** Force curves with an inconspicuous yield point.

The 2D yield strength ($\sigma^{2D}$) can be calculated from:

$$\sigma^{2D} = (FE^{2D}/4\pi R)^{1/2}$$

Where F is the force at the first yield point and R is the tip radius. Sometimes the first yield point may be vague and hard to pick out (**Figure S45b**). However, in such a scenario, a right-shift in the force curve could serve as an indicator that a slight yield happened in the previous force curve.

In **Figure S46**, all data points are plotted in one scheme. The average modulus of **YZ-2** is 50.9 GPa, with a standard deviation of 15.0 GPa. While modulus scatters from 30 to 100 GPa, strength is more condensed from 0.7-1.3 GPa, giving an average of 0.976±0.113 GPa.

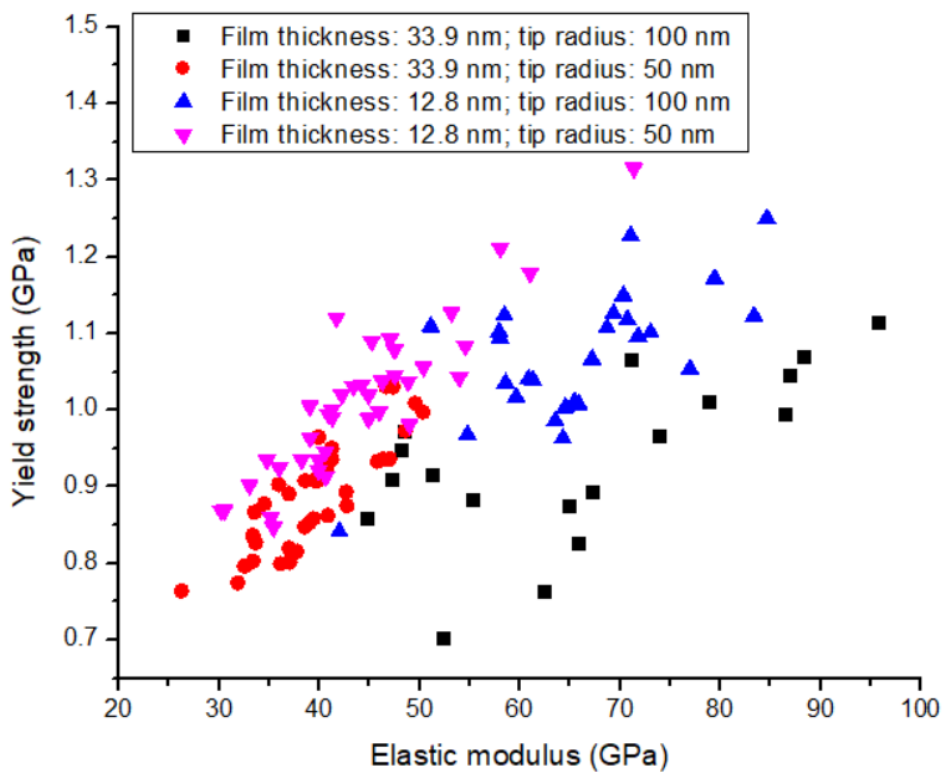

**Figure S46**. A scatter plot of modulus and strength for 120 individual membranes

**Scrolled Fiber Tensile Test**

**Preparation of composite scrolled fibers**: Polycarbonate (PC, Mw: 60K) solutions with different concentrations (1-4% in CHCl$_3$) were spin-coated onto clean SiO$_2$/Si wafers. After annealing at 100 °C, an additional layer of **YZ-2** was introduced by spin-coating and annealing. The resulting composite nanostructures were further scrolled under transverse force to offer desired scrolled fibers (**Figure S50**).[38] All fibers were vacuum dried at 65 °C for 10 h before the test.

**Note**: The wafers we used herein have a uniform size of 3.5*4.5 cm. After spin-coating of the PC film and the **YZ-2** film, the edge of the composite film is trimmed with a razor blade and the film size is 2.8*3.9 cm. In this study, the scrolling is always along the long axis. So, the length of the resulting fiber is around 2.8 cm (gauge length: 1.6 cm).

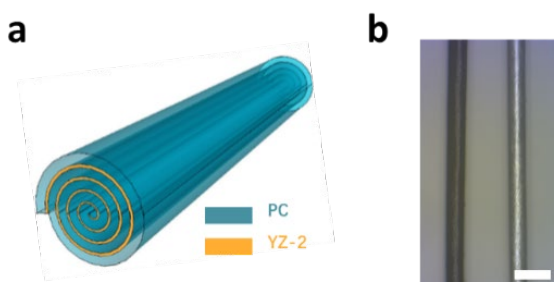

**Figure S50**. **a**, Schematic illustration of a composite scroll fiber. **b**, Micrograph of a human hair (left) and a scroll fiber (right). Scale bar, 100 μm.

**Thickness measurement**: The thickness of PC films was determined by an *XLS-100* ellipsometer from *J.A. Woollam Co.* (**Figure S51**). **YZ-2** films were spin-coated on PC films, then transferred and washed with CHCl$_3$. The thickness of **YZ-2** films was then determined by an AFM at scratches in 8 different places.

Spin-coating condition **I**: 2.5 mg/mL, 1500 rpm, 1min; then 100 °C, 1 min.

The thickness of **YZ-2** film: 8.5±1.37 nm (from 8 different places)

Spin-coating condition **II**: 5.0 mg/mL, 1000 rpm, 1min; then 100 °C, 1 min.

The thickness of **YZ-2** film: 21.1±1.90 nm (from 8 different places)

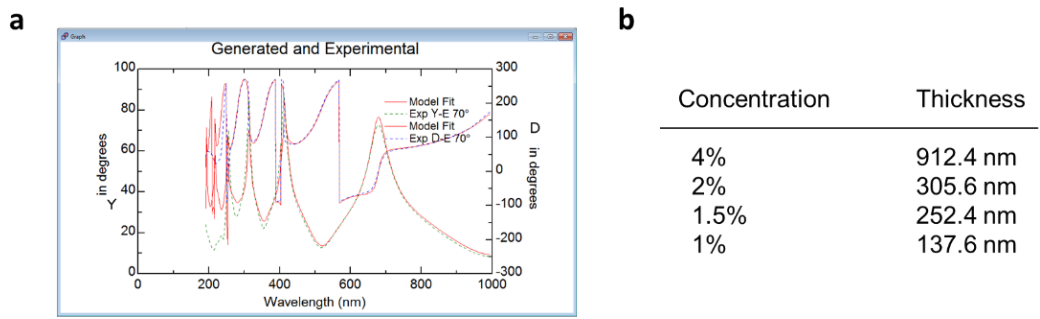

**Figure S51**. **a**, Experimental data from ellipsometer (blue curves) and its thickness fitting (red curves). **b**, Thickness of PC films with different PC concentrations. Spin-coating conditions: 4000 rpm, 1 min; then annealed at 100 °C for 4 min.

The volume fraction ($V_{2DP}$) can be determined as:

$$V_{2DP} = \frac{Thickness_{2DP}}{Thickness_{PC}}$$

**Scroll fiber tensile test**: The tensile test was performed on an *Instron 8848 Micro Tester*. Firstly, the scrolled fiber was glued onto a hollow cardboard using epoxy resin, with a gauge length of 16mm (**Figure S52a**). Then mount the whole sample onto the Instron micro tester, cut the connecting parts on the cardboard and let the scroll fiber free-stand. The test was carried out at room temperature with a strain rate of 0.1 mm/s using a 10-N load cell. The force-displacement curve is recorded until the fiber breaks off (**Figure S52b**).

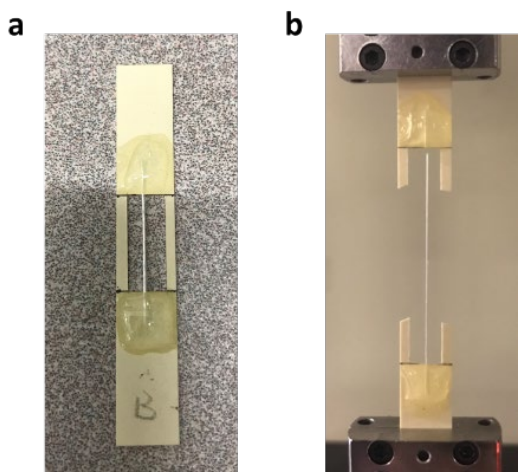

**Figure S52**. **a**, Before the tensile test. **b**, After the tensile test.

**Data analysis**: The engineering strain ($\varepsilon_E$) is calculated from the elongation and the original fiber length (16 mm). Meanwhile, the engineering stress ($\sigma_E$) is obtained by dividing the force by the fiber cross-sectional area, which equals to the thickness of the composite film times its length.

The true strain ($\varepsilon_{tr}$) and the true stress ($\sigma_{tr}$) can be converted from their engineering strain and stress, using the following equations:

$$\varepsilon_{tr} = \ln(1+ \varepsilon_E);$$

$$\sigma_{tr} = \sigma_E(1+ \varepsilon_E)$$

The elastic modulus (*E*) is calculated from the very first part (<2%) of the stress-strain curve, in which the curve is linear. The ultimate tensile strength ($\sigma$) is obtained from the failure point.

By combining different PC and **YZ-2** spin-coating conditions, we prepared and measured composite fibers with five different volume fractions (0.9%, 2.3%, 6.9%, 7.7%, and 13.3%). The results are shown below, along with their PC control fibers (**Figure S53**).

We also compared these **YZ-2**/PC composite scroll fiber results with our previous scroll fiber results from graphene/PC composites (**Figure S54**).[38] The graphene also shows a significant increase in strength even at a very low volume fraction (0.2%), indicating a higher enhancement efficiency. However, it decreases the fiber tensile modulus dramatically. This is attributed to the telescoping effect, which is common in 2D material/polymer composite scroll fibers.[38,39] In the **YZ-2**/PC case, due to the highly polarized surfaces, **YZ-2** molecules are not only binding tightly with each other, but also very "sticky" to the PC film surface. This strong inter-material interaction would be crucial for further composite development.

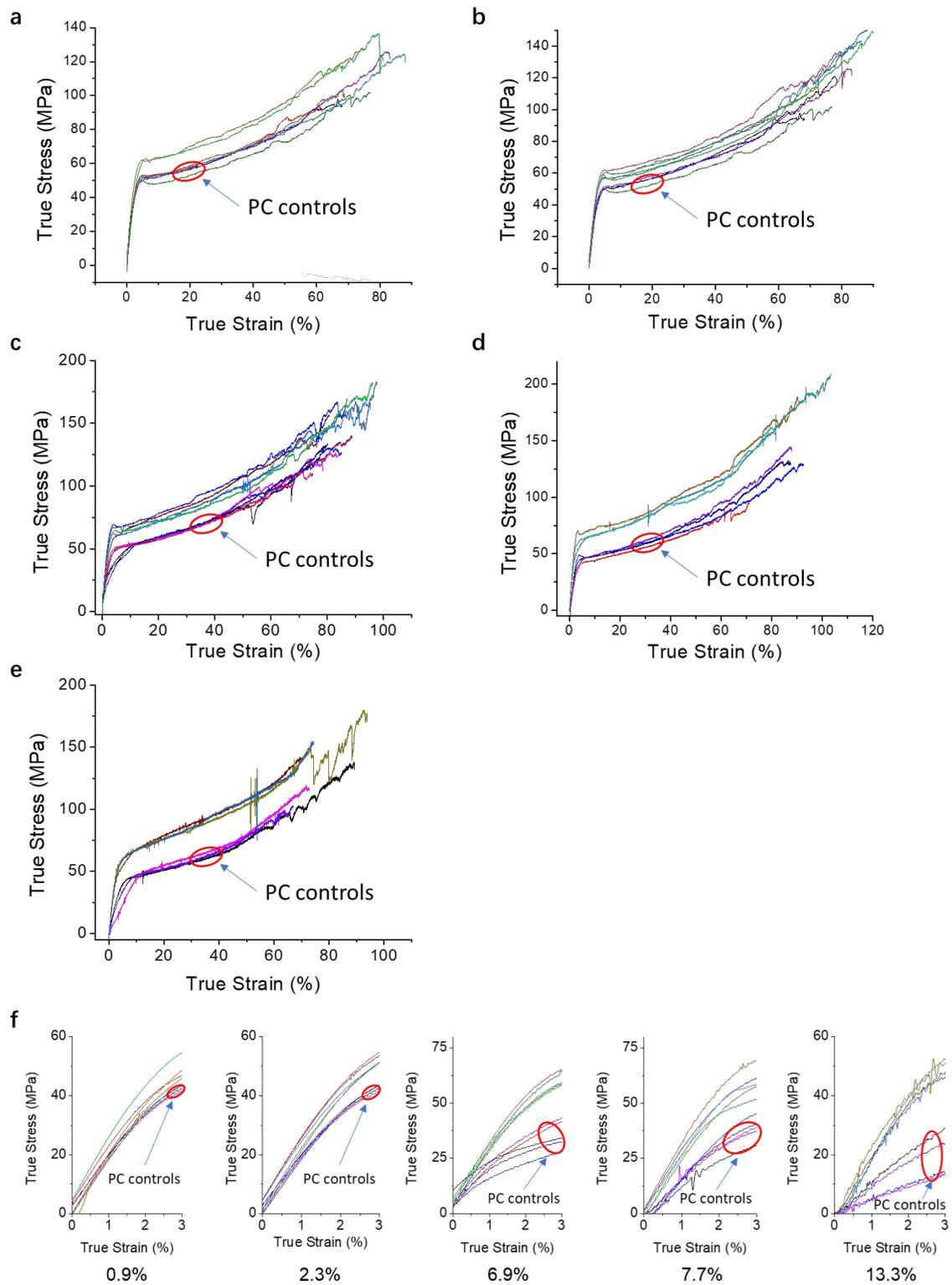

**Figure S53.** True stress-strain plots of composite scroll fibers and their PC controls. **a**, $V_{2DP}$ = 0.9%. **b**, $V_{2DP}$ = 2.3%. **c**, $V_{2DP}$ = 6.9%. **d**, $V_{2DP}$ = 7.7%. **e**, $V_{2DP}$ = 13.3%. **f**, Zoomed-in images for elastic modulus calculation. Strain increment range for linear fitting: 0-2%.

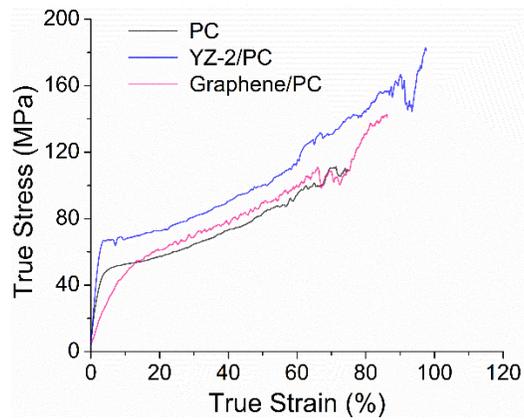

**Figure S54**. A comparison between a **YZ-2**/PC scroll fiber and a graphene/PC scroll fiber, PC control is also shown. Volume fraction of **YZ-2**: 6.9%; volume fraction of graphene: 0.19%. Graphene/PC data is reproduced from ref 15.